\documentclass[twocolumn]{aastex631}
\usepackage{graphicx}
\usepackage{xcolor}
\usepackage{placeins}
\usepackage{needspace}

\newcommand{\Zsun}{\ensuremath{Z_{\odot}}}

\newcommand{\HST}{\textit{HST}}

\newcommand{\JWST}{\textit{JWST}}

\newcommand{\wl}{\mbox{$\lambda$}}
\newcommand{\ionrm}[1]{\mbox{\small\sc{\romannumeral #1}}}
\newcommand{\forb}[2]{\mbox{[#1~\ionrm{#2}]}}
\newcommand{\forbdw}[3]{\mbox{[#1~\ionrm{#2}]~\wl\wl#3}}

\newcommand{\OIIw}{\forbdw{O}{2}{3727}}

\newcommand{\OIII}{\forb{O}{3}}
\newcommand{\OII}{\forb{O}{2}}
\newcommand{\NeIII}{\forb{Ne}{3}}
\newcommand{\Ha}{\mbox{H\ensuremath{\alpha}}}
\newcommand{\Hb}{\mbox{H\ensuremath{\beta}}}

\newcommand{\grizli}{\textsc{grizli}}
\newcommand{\msaexp}{\textsc{msaexp}}

\newcommand{\comment}[1]{}

\usepackage{float}

\begin{document}

\title{High-Redshift Galaxy Candidates at $z = 6+$ as Revealed by \JWST\ Observations of MACS0647}

\newcommand{\CfA}{\affiliation{Center for Astrophysics \text{\textbar} Harvard \& Smithsonian, 60 Garden Street, Cambridge, MA 02138, USA}}

\newcommand{\STScI}{\affiliation{Space Telescope Science Institute (STScI), 3700 San Martin Drive, Baltimore, MD 21218, USA}}

\newcommand{\JHU}{\affiliation{Center for Astrophysical Sciences, Department of Physics and Astronomy, The Johns Hopkins University, 3400 N Charles St. Baltimore, MD 21218, USA}}

\newcommand{\ESAAURA}{\affiliation{Association of Universities for Research in Astronomy (AURA), Inc.~for the European Space Agency (ESA)}}

\newcommand{\TAMUG}{\affiliation{George P. and Cynthia Woods Mitchell Institute for Fundamental Physics and Astronomy, Texas A\&M University, College Station, TX 78743, USA}}

\newcommand{\TAMU}{\affiliation{Department of Physics and Astronomy, Texas A\&M University, 4242 TAMU, College Station, TX 78743, USA}}

\newcommand{\RIT}{\affiliation{School of Physics and Astronomy, Rochester Institute of Technology, 84 Lomb Memorial Drive, Rochester, NY 14623, USA}}

\newcommand{\UT}{\affiliation{Department of Astronomy, The University of Texas at Austin, Austin, TX 78712, USA}}


\correspondingauthor{Keduse Worku}
\email{kworku2@jhu.edu}

\author[0000-0002-2289-5541]{Keduse Worku} \JHU \STScI

\author[0000-0003-4512-8705]{Tiger Yu-Yang Hsiao} 
\affiliation{Cosmic Frontier Center, The University of Texas at Austin, Austin, TX 78712, USA} 
\affiliation{Department of Astronomy, The University of Texas at Austin, Austin, TX 78712, USA}

\author[0000-0001-7410-7669]{Dan Coe} \STScI \ESAAURA \JHU

\author[0000-0002-5258-8761]{Abdurro'uf} 
\affiliation{Department of Astronomy, Indiana University, 727 East Third Street, Bloomington, IN 47405, USA}
\JHU \STScI

\author[0009-0007-0522-7326]{Tom Resseguier} \JHU \STScI

\author[0000-0003-2366-8858]{Rebecca L. Larson} \RIT

\author[0000-0002-0243-6575]{Jacqueline Antwi-Danso}
\affiliation{David A. Dunlap Department of Astronomy and Astrophysics, University of Toronto, 50 St. George Street, Toronto, Ontario, M5S 3H4, Canada}
\altaffiliation{Banting Postdoctoral Fellow}

\author[0000-0003-2680-005X]{Gabriel Brammer}
\affiliation{Cosmic Dawn Center (DAWN), Copenhagen, Denmark}
\affiliation{Niels Bohr Institute, University of Copenhagen, Jagtvej 128, Copenhagen, Denmark}

\author[0000-0002-5588-9156]{Vasily Kokorev} \UT 

\author[0000-0002-7908-9284]{Larry D. Bradley} \STScI

\author[0000-0001-6278-032X]{Lukas J. Furtak}
\affiliation{Cosmic Frontier Center, The University of Texas at Austin, Austin, TX 78712, USA} 
\affiliation{Department of Astronomy, The University of Texas at Austin, Austin, TX 78712, USA}

\author[0000-0003-3484-399X]{Masamune Oguri}
\affiliation{Center for Frontier Science, Chiba University, 1-33 Yayoi-cho, Inage-ku, Chiba 263-8522, Japan}
\affiliation{Department of Physics, Graduate School of Science, Chiba University, 1-33 Yayoi-Cho, Inage-Ku, Chiba 263-8522, Japan}




\begin{abstract}

We present a catalog of 57 high-redshift $z>6$ galaxy candidates, including 14 spectroscopic confirmations ($z = 6.10$ -- 9.25), 2 Little Red Dots ($z = 4.77$, 5.81), and 2 interlopers ($z = 3.23$, 3.72), based on \JWST\ NIRCam imaging (7 filters), NIRSpec spectroscopy (PRISM and G395H), and archival \HST\ imaging (17 filters) of the strong lensing galaxy cluster MACS0647. Our highest redshift confirmation ($z = 9.25$) is an Extremely Blue Galaxy (presented in~\citealt{Yanagisawa2024}), and here we identify a spectral turnover likely due to damped Lyman-$\alpha$. We identify an overdensity of galaxies with spectroscopic redshifts $z = 6.1$, confirming the $z \sim 6$ overdensity identified in \HST\ images. In one of these galaxies, our high-resolution G395H spectroscopy reveals two spatially resolved components with a velocity difference of $\sim$90 km/s; if these components are gravitationally bound, this would imply a dynamical mass on the order of $\sim 10^8\ M_\odot$ given their projected separation. We present spectral line fluxes, widths, and derived physical properties, including stellar masses ($10^8 - 10^9  \ \mathrm{M}_{\odot}$) and metallicities ($10\% - 40\% \ \mathrm{Z}_{\odot}$) for our spectroscopic sample. We note half of our NIRSpec data was obtained with standard 3-slitlet nods and half was obtained with single slitlets yielding similar results, demonstrating the power to observe more sources on a densely packed NIRSpec MSA.

\end{abstract}



\section{Introduction} \label{sec:intro}
A significant phase transition occurred in the Intergalactic Medium (IGM) as it shifted from a neutral to an ionized state at $z \sim 6$ (about 1 Gyr after the Big Bang), marking the period known as the Epoch of Reionization \citep[EoR;][]{Loeb2001}. The culprits behind this reionization phase are widely believed to be the intense UV radiation emitted by the first generation of galaxies and stars \citep[e.g.,][]{Fan2006, Bouwens2012, Duncan2015, dayal2018, Naidu2020}. The crucial task of detecting and characterizing galaxies at high redshifts becomes paramount in comprehending the evolutionary trajectory of galaxies, as well as in gaining insight into the mechanisms behind the EoR \citep[e.g.,][]{Robertson2022}. However, our ability to study early galaxies remains constrained by the limitations in telescope sensitivity.

Thanks to the advanced capabilities of the {\em JWST} \citep{Gardner2006} a multitude of galaxies with redshifts $> 10$ have been brought to light through photometric discovery utilizing NIRCam \citep{Rigby2022, Harikane2022, Castellano2022, Bradley2022, Naidu2022, Finkelstein2022, Yan2023, Adams2023, Whitler2023, Donnan2023, Atek2023, Harikane2023, Hsiao2023a, Casey2023}. This impressive feat has been complemented by spectroscopic confirmations facilitated by NIRSpec \citep[e.g.,][]{Kokorev2025, Perez-Gonzalez2025, Castellano2025, CurtisLake2023, Wang2023, Hsiao2023b, Goulding2023}, including the identification of two compelling galaxies with redshifts surpassing 13~\citep{Naidu2025, Wu2025, Carniani2024, CurtisLake2023, Wang2023}. This collection of unparalleled data has granted us the opportunity to delve into an array of studies encompassing luminosity functions \citep[e.g.,][]{Donnan2024, Donnan2023, Harikane2023}, stellar masses \citep[e.g.,][]{Hsiao2023a, Mowla2024}, star-formation histories \citep[e.g.,][]{Hsiao2023a}, cosmic star formation history \citep[e.g.,][]{Harikane2023}, as well as Lyman-alpha damping wings \citep[e.g.,][]{Hsiao2023b, Heintz2023, Heintz2025} within the context of the high-redshift universe.

Gravitational lensing induced by massive galaxy clusters serves to magnify both the light emitted by distant objects and their physical dimensions. This cosmic telescope not only brings the fluxes of faint objects from the early Universe into the observable range, but also enhances the visibility of small-scale structures \citep[e.g.,][]{Coe2013, Welch2022, Vanzella2022, Mestric2022, Welch22_Earendel, Welch2022JWST, Claeyssens2023, Hsiao2023a, Fujimoto2024, Mowla2024, Adamo2024, Nightingale2025}. Consequently, gravitational lensing has enabled us to unveil and scrutinize early galaxies along with their intrinsic properties. In pursuit of investigating numerous pivotal scientific inquiries pertaining to the early Universe, a variety of surveys centered around lensing clusters have been undertaken. These initiatives encompass the Cluster Lensing and Supernova survey with Hubble\footnote{\url{https://www.stsci.edu/~postman/CLASH/}} \citep[CLASH;][]{Postman2012}, the Hubble Frontier Fields project \citep{Lotz2017}, and the Reionization Lensing Cluster Survey\footnote{\url{https://relics.stsci.edu}} \citep[RELICS;][]{Coe2019}.

CLASH stands as one of the significant Hubble treasury programs that has embraced the lensing technique for the exploration of distant galaxies \citep[e.g.,][]{Zheng2012, Coe2013, Bouwens2014, Smit2014, Bradley2014}. This initiative captured imagery of 25 massive galaxy clusters across 16 filters using the {\HST}, spanning a wavelength range from the near-ultraviolet ($\sim200\,{\rm nm}$) to the near-infrared ($\sim1.6\,{\rm \mu m}$), with inclusion of the galaxy cluster MACSJ0647.7+7015 (MACS0647; $z=0.591$) \citep{Ebeling2007} modeled by \citep{Zitrin2011}. Through CLASH observations, \citet{Bradley2014} identified 32 lensed candidates at redshifts around $z \sim 6$ to 8. In addition, a triply-lensed candidate with a redshift of approximately $z\sim11$, named MACS0647--JD, was discovered in the same field \citep{Coe2013}. Furthermore, {\em JWST}/NIRCam observations unveiled the intricate structure of MACS0647--JD, resolving it into two components down to a scale of 20$\,$pc \citep{Hsiao2023a}. Subsequent {\em JWST}/NIRSpec observations confirmed the redshift of MACS0647--JD to be $z=10.17$ and revealed seven emission lines \citep{Hsiao2023b}. In particular, {\em JWST} observations of MACS0647 also identified two potential lensed stars with a redshift of $z=4.76$ \citep{Meena2023,Furtak2024}.

Despite the numerous high-z candidates unveiled by {\em JWST}, a peculiar case of misclassification involving the galaxy CEERS-93316 emerges. This galaxy, initially assigned a photometric redshift of $z=16$ \citep{Donnan2023}, has since been validated at a spectroscopic redshift of $z=4.9$ as indicated by \citet{ArrabalHaro2023} due to its pronounced emission lines. In particular, galaxies with robust optical nebular emission lines at $z<6$ can mimic the NIRCam photometric profiles of $z > 7 - 10$ Lyman Break Galaxies \citep[e.g.,][]{McKinney2023}.  Furthermore, the issue of dusty galaxies has long posed a significant challenge in tainting high redshift galaxy observations within photometric data, a predicament predating the era of {\em JWST} \citep[e.g.,][]{Naidu2022dusty}. Thus, the confirmation of redshifts demands complementary spectroscopic follow-up investigations. This potential quandary may perplex astronomers as the era of {\em JWST}, {\em Euclid}, and the Roman Space Telescope yields unparalleled photometric data, yet remains constrained by a scarcity of spectroscopic data~\citep{Rigby2022, Mellier2025_EuclidOverview, Roman_2021}.
\begin{figure*}
\centering
\includegraphics[width=0.498\textwidth]{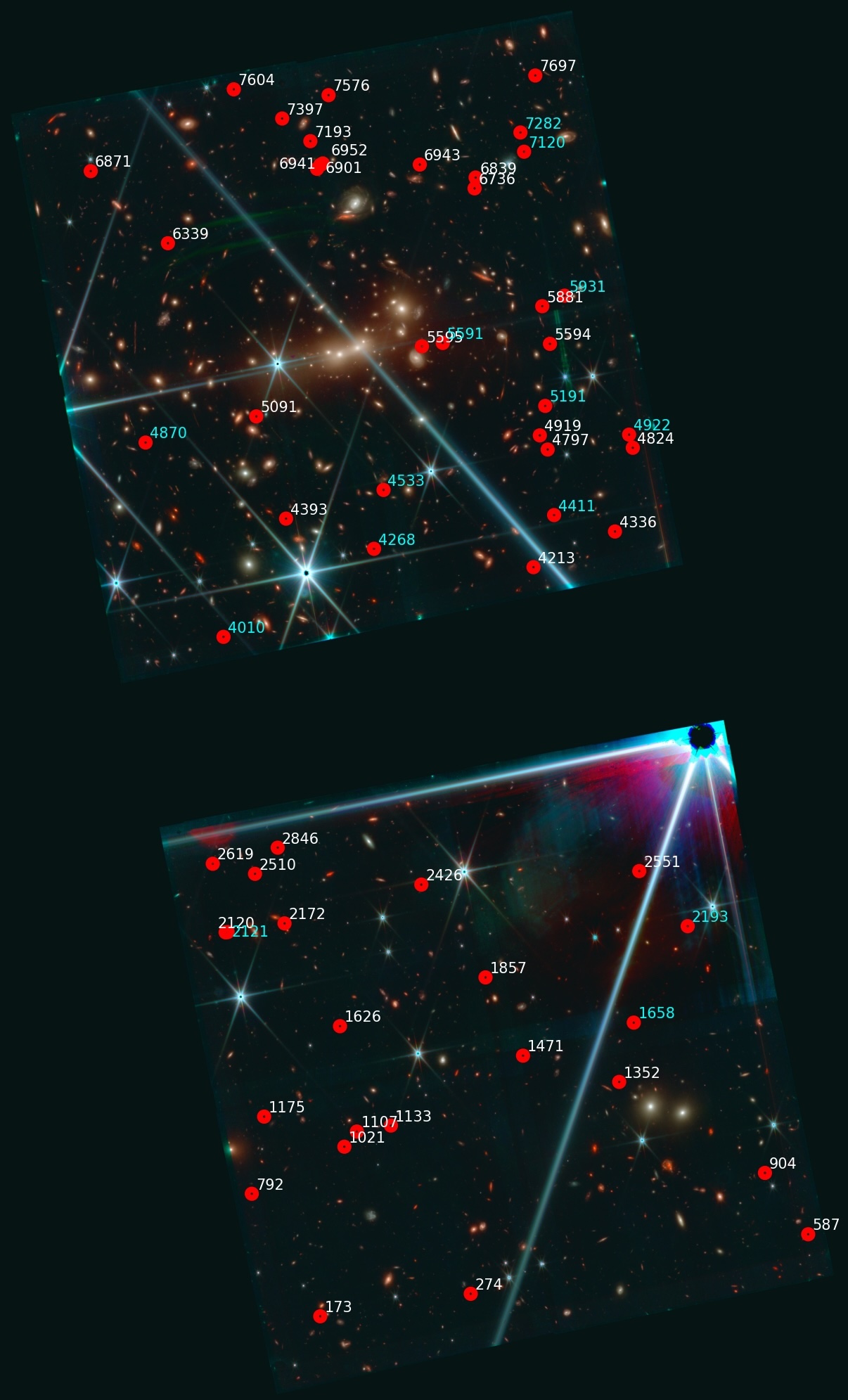}
\includegraphics[width=0.492\textwidth]{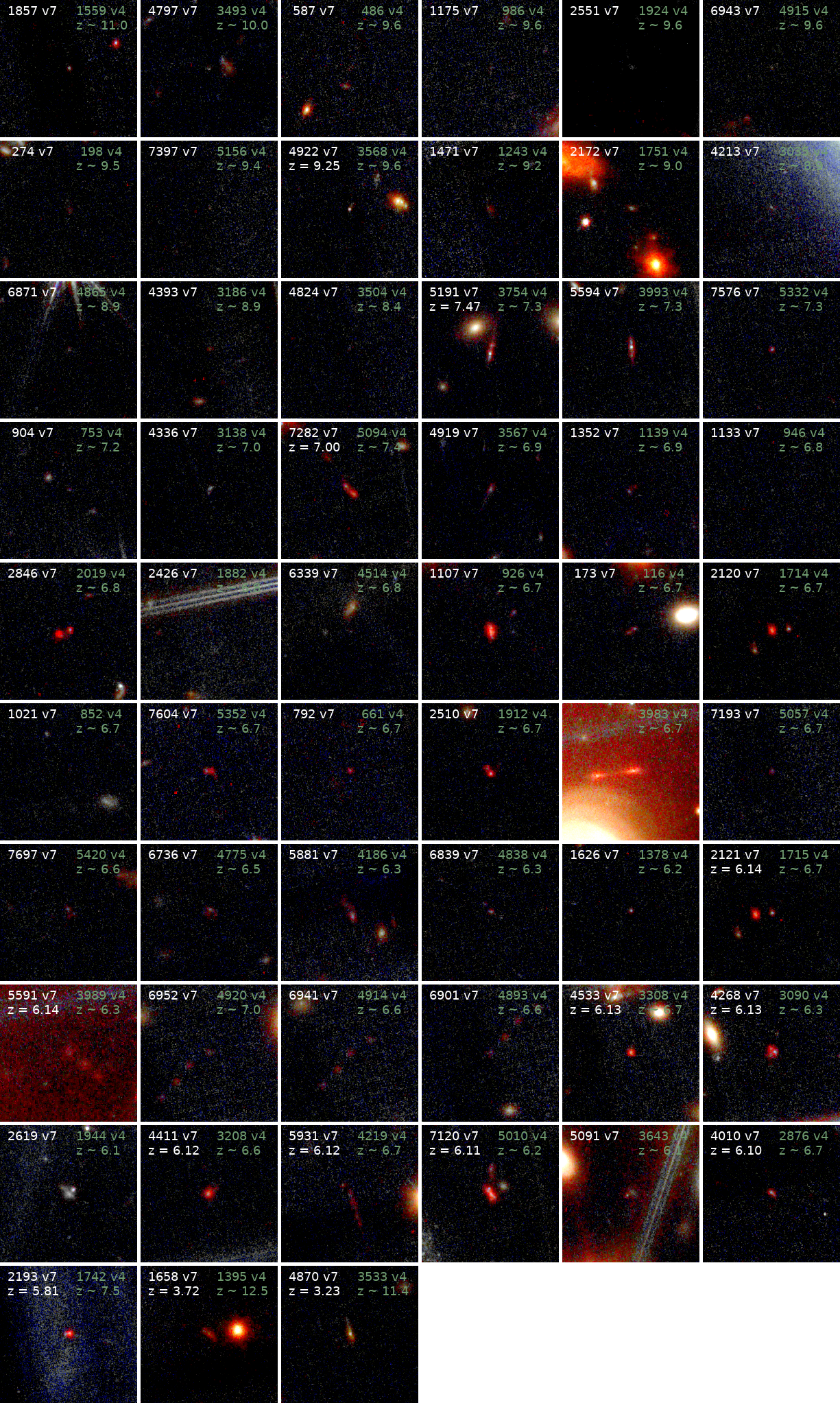}
\caption{\textbf{Left:} Overview MACS0647 NIRCam color image based on DJA v7 reductions of observations in six wide filters. High-redshift candidates are labeled in white (v7 ID numbers), while galaxies with spectroscopic redshifts are indicated in cyan. \textbf{Right:} \JWST\ NIRCam $4\arcsec \times 4\arcsec$ color images, each centered on a high-redshift candidate in the final sample.
}
\label{fig:labeled_color_image}
\end{figure*}

In this work, we present and analyze a sample of approximately 50 galaxy candidates at a redshift of $z>6$ in the recent observations of MACS0647 using 7 \JWST\ NIRCam filters (1.15 -- 5$\,\rm{\mu m}$). We investigate their early galaxy evolution through detailed physical characterizations, such as stellar masses and dust content, while constraining their star formation history. Furthermore, we spectroscopically confirmed 15 of these galaxies through JWST NIRSpec data with Hi-Res and prism modes (including MACS0647--JD; see also \citealt{Hsiao2023b, Abdurrouf2024}), additionally revealing two low-redshift interlopers. Candidates of note include strong $\mathrm{Lyman-\alpha}$ emitters, a Little Red Dot (LRD), and an overdensity of galaxies at $z=6.1$, indicating possible cluster islands of reionization. 

This paper is organized as follows: Section \S\ref{sec:data} details the {\em JWST} and {\em HST} observational datasets and data reduction; Section \S\ref{sec:methods} describes our methodology, including our photometric and spectroscopic analysis with high-redshift galaxy selection detailed in \S\ref{sec:candidateselection}; we delve into a comprehensive discussion of our findings in Section \S\ref{sec:results&discussion}, which encompass measurements of physical parameters derived from SED and spectroscopic fitting; finally, Section \S\ref{sec:conclusion} provides our concluding remarks. 

\begin{figure*}[htbp]
\centering

\begin{minipage}[b]{0.14\hsize}
  \raggedright
  v4 ID 3754\\
  v7 ID 5191\\
  $z = 7.46$\\[2pt]
  \includegraphics[width=\hsize]{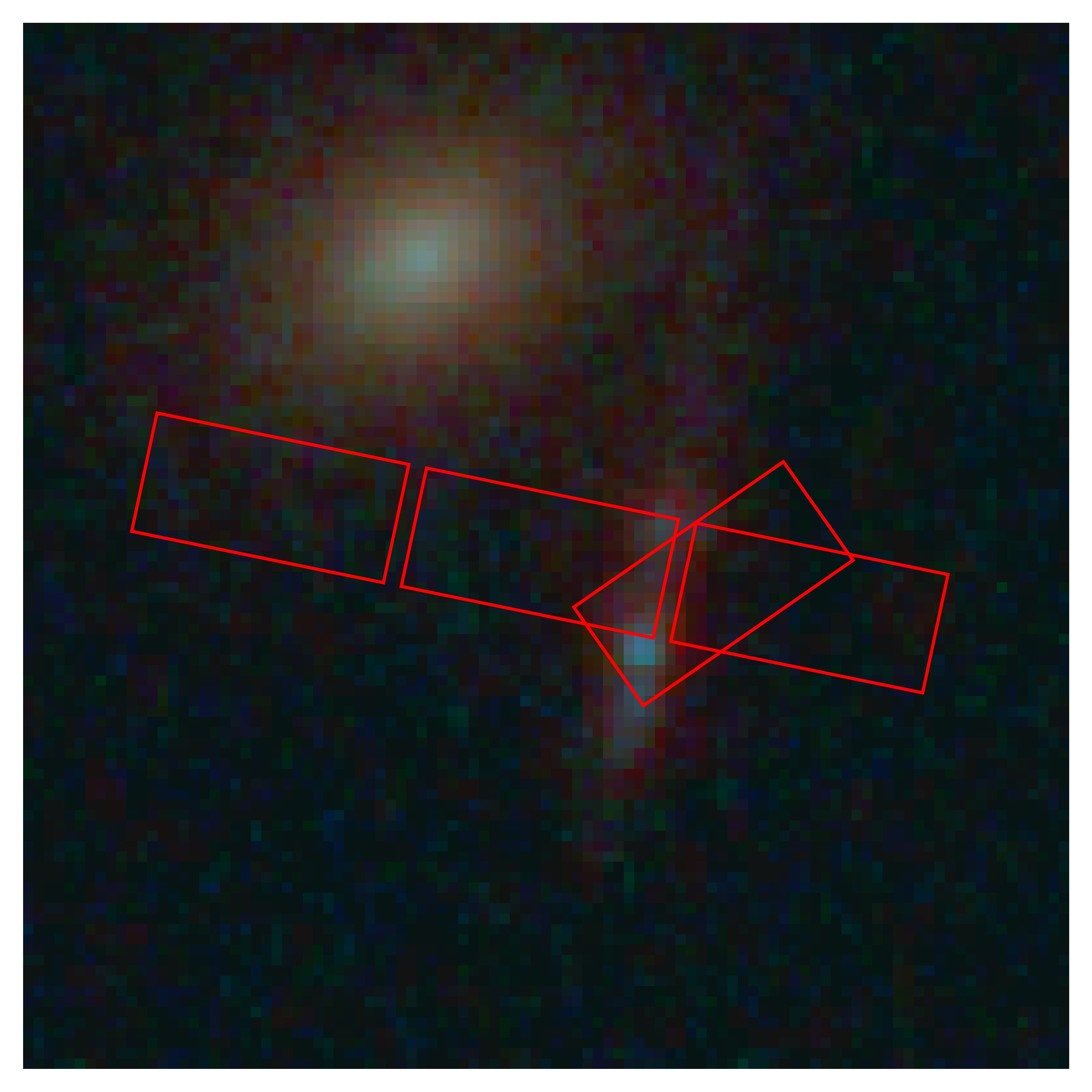}
\end{minipage}
\hfill
\includegraphics[width=0.50\hsize]{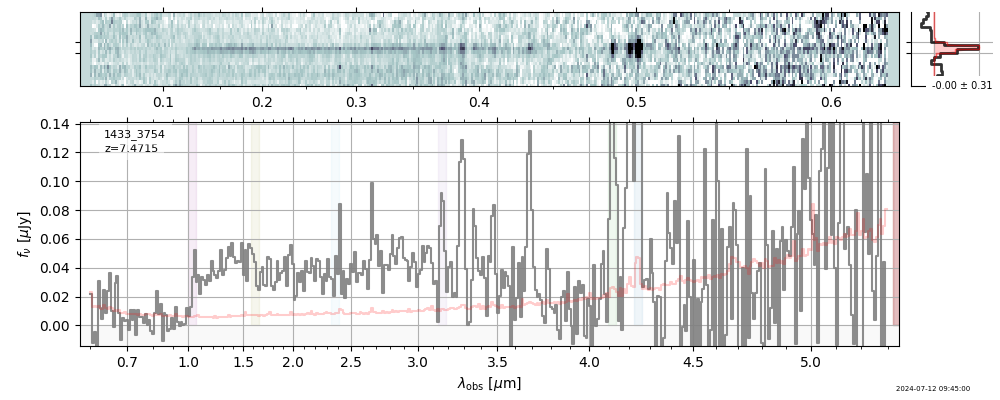}
\hfill
\includegraphics[width=0.33\hsize]{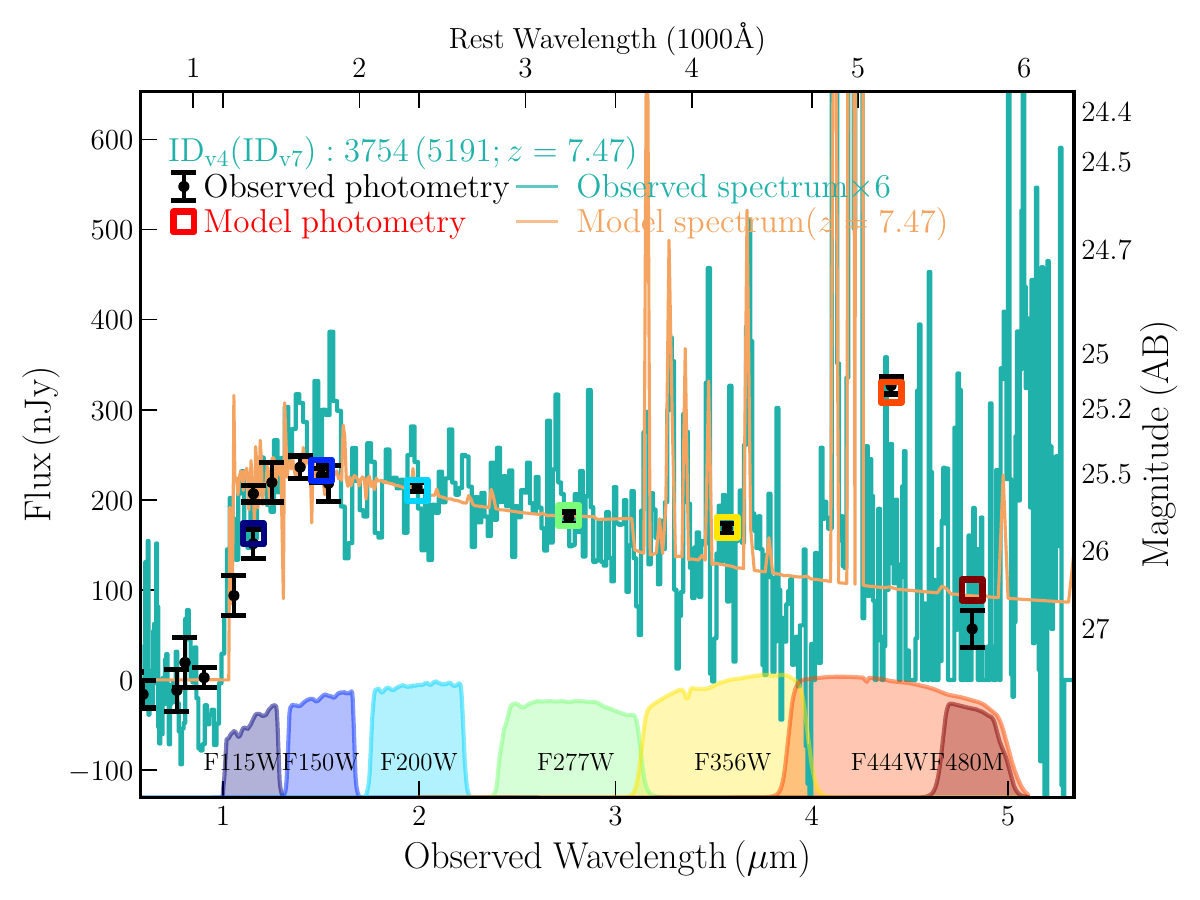}

\caption{
Example of a high-redshift galaxy (v4 3754 / v7 5191, $z = 7.464$) observed with NIRSpec PRISM in both Cycle~1 slitlet configurations. 
This target was observed in Obs~21 using single-slitlet placements and in Obs~23 using the standard 3-slitlet nod pattern (left). 
The PRISM spectrum (middle) shows multiple strong emission lines, and the corresponding SED fit is shown on the right. 
}
\label{fig:3754_prism_sed}
\end{figure*}

\section{Data}
\label{sec:data}
We utilize NIRCam imaging and NIRSpec prism spectroscopy data from the {\em JWST} Cycle 1 program GO 1433 (PI Coe; \citealt{Hsiao2023a}) and NIRSpec G395H high-resolution spectra from the {\em JWST} Cycle 2 program GO 4246 (PI Abdurro'uf; \citealt{Abdurrouf2024}). We also combine the imaging data with the archival {\em HST} imaging from programs GO 9722, 10493, 10793, 12101, and 13317. A thorough description of these data sets is given in \citet{Hsiao2023a, Hsiao2023b, Abdurrouf2024}. In the following, we will only briefly describe some of the data that are used in the analysis of this paper. 

\subsection{\HST\ imaging}
\label{sec:hst}
MACS0647+70 was observed through 39 \HST\ orbits across 17 different filters.
Its initial observations were carried out as part of programs GO 9722 (PI Ebeling), and GO 10493 and 10793 (PI Gal-Yam), employing the ACS F555W and F814W filters. Subsequently, the CLASH program (GO 12101, PI Postman) extended the observations by employing 15 additional filters spanning a wavelength range from 0.2 to 1.7 micrometers, utilizing the WFC3/UVIS, ACS, and WFC3/IR instruments. Furthermore, supplementary imaging in the WFC3/IR F140W filter was acquired as part of a grism spectroscopy program (GO 13317, PI Coe).

\subsection{JWST NIRCam imaging}
\label{sec:nircam}
\JWST\ NIRCam imaging was obtained using 6 filters: F115W, F150W, F200W, F277W, F356W, and F444W, spanning the wavelength range of $1-5\,\micron$. Subsequently, additional NIRCam imaging was obtained in the F200W and F480M filters, concurrently with the NIRSpec observation 21 (Obs 21), as detailed in \S\ref{sec:nirspec}. Each NIRCam image involved a 35-minute exposure time, distributed across 4 \textsc{INTRAMODULEBOX} dithers to cover short-wavelength gaps. For an in-depth understanding of the reduction process, we direct interested readers to consult \citet{Hsiao2023a,Hsiao2023b}. In brief, the NIRCam images underwent processing through the \grizli\ pipeline \citep{Brammer2022}. Initially, \grizli\ v4 images were employed for high-redshift candidate selection (see \S\ref{sec:candidateselection}). Updated zeropoint corrections, based on refined calibrations, were implemented in \grizli\ v5. In this study, we conducted our analysis using \grizli\ v7 image mosaics, which featured improved sky flats, bad pixel tables, wisp templates, and long-wavelength zeropoints. 

Figure~\ref{fig:labeled_color_image} shows the NIRCam color composite of MACS0647 used to identify and label the high-redshift candidates in this work, alongside each individual galaxy candidate.

\subsection{JWST NIRSpec spectroscopy}
\label{sec:nirspec}
The \JWST\ Cycle 1 (GO 1433) NIRSpec multi-object spectroscopy (MOS) observation using the microshutter assembly (MSA)  was conducted on 2023-01-08 (Obs 21) and 2023-02-20 (Obs 23). Both used the low resolution $R\sim 30-300$ prism that covers a wide wavelength range from $0.6-5.3\,\micron$. Obs 21 was performed with single slitlets (plus two dithers), while Obs 23 was performed with standard 3-slitlet nods.  The two observations have a similar exposure time of 1.8 hours. Twenty high redshift candidates (see \S\ref{sec:candidateselection}), including MACS0647$-$JD \citep{Hsiao2023b}, were selected to be observed with NIRSpec based on their photometric redshifts (see also Table \ref{tab:catalog_spec}), with the spectra of ten candidates allowing us to determine their spectroscopic redshift.

In \JWST\ Cycle 2 (GO 4246), we obtained NIRSpec MOS G395H (2.9 -- 5.3$\mu$m) high-resolution ($R \sim 3000$) spectra on 2024-01-14. The standard 3-slitlet nods had a total exposure time of 2.5 hours. Representative G395H spectra are shown in Appendix~A (Figures~\ref{fig:3754_g395h} and \ref{fig:5494_g395h}).

Our NIRSpec data reduction is described in \citep{Abdurrouf2024, Hsiao2024_CO}. Briefly, we retrieved NIRSpec Level 1 data products from MAST and processed them with the STScI \JWST\ pipeline\footnote{https://github.com/spacetelescope/jwst}  and {\tt msaexp}.\footnote{https://github.com/gbrammer/msaexp} We retrieved reduced spectra from the Dawn JWST Archive (DJA)\footnote{\url{https://dawn-cph.github.io/dja/}} and also reprocessed some separately.

For the PRISM data, we find the single-slitlet and 3-slitlet data yield similar results. Our NIRSpec PRISM data is unique in this respect: for some galaxies, we have both observations and can compare directly. For the single-slitlet data, we subtract backgrounds measured in nearby empty slits (added for this purpose). Single-slitlet observations allow more science targets to be densely packed on the NIRSpec MSA. An example galaxy observed with both single-slitlet and 3-slitlet PRISM configurations is shown in Figure~\ref{fig:3754_prism_sed}.

\begin{table*}[htbp]
\centering
\setlength{\tabcolsep}{8pt}
\caption{
Physical properties derived with \textsc{bagpipes} for a subset of our high-redshift candidates, based on NIRCam photometry and multi-band SED fitting. \textbf{Note:} Little Red Dots (LRDs) are marked with an asterisk ($^{*}$); their SED fits may be unreliable due to extreme colors or AGN contamination.} \label{tab:catalog}
\begin{tabular}{lccccccc}
\hline
ID (v7) & ID (v4) & $z_{\rm phot}$ & Stellar Mass & SFR & sSFR & $A_{V}$ & Mass-weighted Age \\
        &         &                & $\log(M_{*}/M_\odot)$ & $(M_{*}/\mathrm{yr})$ & $({\rm yr}^{-1})$ & (mag) & (Myr) \\
\hline
173 & 116 & $6.30^{+0.52}_{-5.14}$ & $8.58^{+0.25}_{-0.94}$ & $2.74^{+2.00}_{-2.67}$ & $-8.27^{+0.30}_{-0.54}$ & $0.23^{+0.28}_{-0.14}$ & $162^{+430}_{-85}$ \\
1107 & 926 & $6.58^{+0.15}_{-0.12}$ & $9.12^{+0.14}_{-0.14}$ & $49.28^{+5.22}_{-3.81}$ & $-7.40^{+0.12}_{-0.16}$ & $0.51^{+0.04}_{-0.03}$ & $ 17^{+ 9}_{- 5}$ \\
1352 & 1139 & $6.93^{+0.74}_{-4.78}$ & $8.54^{+0.17}_{-0.62}$ & $1.94^{+1.49}_{-1.80}$ & $-8.31^{+0.24}_{-0.25}$ & $0.32^{+0.23}_{-0.15}$ & $177^{+139}_{-79}$ \\
2120 & 1714 & $6.69^{+0.03}_{-0.04}$ & $8.88^{+0.06}_{-0.09}$ & $24.93^{+3.85}_{-2.74}$ & $-7.47^{+0.09}_{-0.10}$ & $0.54^{+0.04}_{-0.04}$ & $ 21^{+ 6}_{- 5}$ \\
2193$^{*}$ & 1742 & $7.19^{+0.08}_{-0.09}$ & $10.04^{+0.04}_{-0.03}$ & $39.10^{+5.99}_{-15.60}$ & $-8.44^{+0.06}_{-0.27}$ & $0.70^{+0.05}_{-0.14}$ & $240^{+56}_{-32}$ \\
2510 & 1912 & $6.72^{+0.05}_{-0.08}$ & $8.99^{+0.13}_{-0.14}$ & $13.13^{+3.13}_{-3.74}$ & $-7.87^{+0.22}_{-0.27}$ & $0.43^{+0.06}_{-0.07}$ & $ 58^{+58}_{-25}$ \\
2619 & 1944 & $6.81^{+0.03}_{-0.03}$ & $9.90^{+0.03}_{-0.03}$ & $29.45^{+3.28}_{-2.00}$ & $-8.44^{+0.07}_{-0.04}$ & $0.41^{+0.03}_{-0.03}$ & $239^{+23}_{-39}$ \\
2846 & 2019 & $6.49^{+0.36}_{-0.19}$ & $9.08^{+0.11}_{-0.11}$ & $5.26^{+2.14}_{-1.03}$ & $-8.38^{+0.18}_{-0.10}$ & $0.36^{+0.12}_{-0.08}$ & $207^{+56}_{-72}$ \\
4010 & 2876 & $6.97^{+0.19}_{-0.17}$ & $8.90^{+0.08}_{-0.12}$ & $3.90^{+1.28}_{-0.82}$ & $-8.32^{+0.21}_{-0.11}$ & $0.13^{+0.09}_{-0.07}$ & $184^{+51}_{-76}$ \\
4213 & 3035 & $7.67^{+0.40}_{-0.31}$ & $7.82^{+0.26}_{-0.24}$ & $1.61^{+0.41}_{-0.33}$ & $-7.61^{+0.30}_{-0.34}$ & $0.02^{+0.04}_{-0.02}$ & $ 30^{+42}_{-17}$ \\
4336 & 3138 & $7.62^{+0.13}_{-0.13}$ & $8.09^{+0.11}_{-0.11}$ & $2.88^{+0.18}_{-0.20}$ & $-7.62^{+0.11}_{-0.15}$ & $0.01^{+0.01}_{-0.01}$ & $ 31^{+14}_{- 8}$ \\
4411 & 3208 & $6.61^{+0.01}_{-0.02}$ & $9.06^{+0.06}_{-0.06}$ & $28.03^{+4.15}_{-4.08}$ & $-7.61^{+0.10}_{-0.10}$ & $0.36^{+0.03}_{-0.03}$ & $ 30^{+ 9}_{- 7}$ \\
4797 & 3493 & $9.95^{+0.40}_{-0.32}$ & $8.17^{+0.22}_{-0.25}$ & $2.05^{+0.70}_{-0.41}$ & $-7.86^{+0.31}_{-0.27}$ & $0.06^{+0.10}_{-0.04}$ & $ 58^{+53}_{-32}$ \\
4922 & 3568 & $9.67^{+0.23}_{-0.23}$ & $8.11^{+0.17}_{-0.15}$ & $2.59^{+0.49}_{-0.52}$ & $-7.70^{+0.20}_{-0.22}$ & $0.01^{+0.02}_{-0.01}$ & $ 38^{+27}_{-16}$ \\
5191 & 3754 & $7.68^{+0.11}_{-0.10}$ & $8.37^{+0.06}_{-0.05}$ & $23.70^{+3.59}_{-2.72}$ & $-7.00^{+0.00}_{-0.00}$ & $0.11^{+0.03}_{-0.04}$ & $  2^{+ 0}_{- 0}$ \\
5591 & 3989 & $6.31^{+0.22}_{-4.39}$ & $8.07^{+0.18}_{-0.49}$ & $0.68^{+0.47}_{-0.57}$ & $-8.30^{+0.31}_{-0.23}$ & $0.30^{+0.36}_{-0.19}$ & $172^{+122}_{-92}$ \\
5881 & 4186 & $6.76^{+0.27}_{-0.11}$ & $8.73^{+0.13}_{-0.14}$ & $7.73^{+1.20}_{-0.97}$ & $-7.84^{+0.19}_{-0.18}$ & $0.02^{+0.03}_{-0.01}$ & $ 55^{+30}_{-22}$ \\
5931 & 4219 & $6.77^{+0.50}_{-0.31}$ & $7.52^{+0.18}_{-0.26}$ & $0.23^{+0.09}_{-0.06}$ & $-8.17^{+0.33}_{-0.23}$ & $0.10^{+0.13}_{-0.07}$ & $127^{+96}_{-72}$ \\
6736 & 4775 & $6.93^{+0.17}_{-0.19}$ & $8.71^{+0.09}_{-0.12}$ & $2.91^{+0.71}_{-0.41}$ & $-8.24^{+0.20}_{-0.15}$ & $0.05^{+0.06}_{-0.04}$ & $151^{+63}_{-60}$ \\
6839 & 4838 & $6.69^{+0.16}_{-0.09}$ & $8.72^{+0.11}_{-0.16}$ & $3.09^{+0.91}_{-0.46}$ & $-8.21^{+0.22}_{-0.18}$ & $0.05^{+0.06}_{-0.04}$ & $139^{+77}_{-59}$ \\
6901 & 4893 & $6.81^{+0.29}_{-0.21}$ & $8.62^{+0.11}_{-0.16}$ & $2.18^{+0.95}_{-0.55}$ & $-8.27^{+0.23}_{-0.15}$ & $0.15^{+0.12}_{-0.09}$ & $162^{+71}_{-70}$ \\
6941 & 4914 & $6.88^{+0.50}_{-5.57}$ & $8.60^{+0.14}_{-0.92}$ & $2.47^{+2.02}_{-2.43}$ & $-8.29^{+0.31}_{-0.63}$ & $0.40^{+0.60}_{-0.17}$ & $167^{+604}_{-90}$ \\
7120 & 5010 & $6.55^{+0.02}_{-0.02}$ & $8.98^{+0.08}_{-0.10}$ & $38.57^{+8.65}_{-3.64}$ & $-7.37^{+0.07}_{-0.08}$ & $0.32^{+0.05}_{-0.03}$ & $ 16^{+ 4}_{- 3}$ \\
7282 & 5094 & $7.21^{+0.12}_{-0.12}$ & $8.78^{+0.08}_{-0.14}$ & $7.00^{+1.98}_{-1.76}$ & $-7.93^{+0.24}_{-0.19}$ & $0.16^{+0.05}_{-0.06}$ & $ 67^{+41}_{-31}$ \\
7576 & 5332 & $7.28^{+0.13}_{-0.12}$ & $7.50^{+0.26}_{-0.21}$ & $2.38^{+0.39}_{-0.45}$ & $-7.06^{+0.06}_{-0.32}$ & $0.03^{+0.04}_{-0.02}$ & $  6^{+10}_{- 4}$ \\
7604 & 5352 & $6.48^{+0.04}_{-0.07}$ & $8.36^{+0.18}_{-0.20}$ & $7.99^{+1.14}_{-1.08}$ & $-7.44^{+0.19}_{-0.22}$ & $0.29^{+0.06}_{-0.04}$ & $ 19^{+14}_{- 8}$ \\
7697 & 5420 & $7.08^{+0.23}_{-0.19}$ & $8.61^{+0.09}_{-0.16}$ & $2.37^{+0.80}_{-0.45}$ & $-8.24^{+0.24}_{-0.16}$ & $0.08^{+0.07}_{-0.05}$ & $147^{+72}_{-67}$ \\
\hline
\end{tabular}
\end{table*}

\section{Methods} 
\label{sec:methods}

\subsection{Candidate Selection}
\label{sec:candidateselection}
To select candidates $z \geq 6$ from our sample, we use a combination of photometric redshift fitting using EAZY (described in Sec.~\ref{sec:sedfitting}) and significance of detection in band signals. The simultaneous analysis of all photometric bands in SED fitting and its extensive use in the literature allows for effective high-redshift candidate selection \citep{Bradley2022, Finkelstein2022, Naidu2022c, Adams2023, Donnan2023, Finkelstein2023}.

Our primary selection criteria are: 
\begin{enumerate}
    \item Signal-to-noise ratio $\geq$ 3 in the F356W band
    \item Signal-to-noise ratio $\geq$ 3 in at least two other filters
    \item Best-fit photometric redshift z $\geq 6$
\end{enumerate}
 
The signal-to-noise criteria across multiple photometric bands supports the robustness of candidate detections. Additionally, each filter image and best-fit SED was reviewed to eliminate diffraction spikes and instrument artifacts.
\begin{figure*} [htbp] 

\begin{minipage}[b]{0.12\hsize}
  \raggedright
  v4 ID 3989\\
  v7 ID 5591\\
  $z = 6.15$
  \includegraphics[width=\hsize]{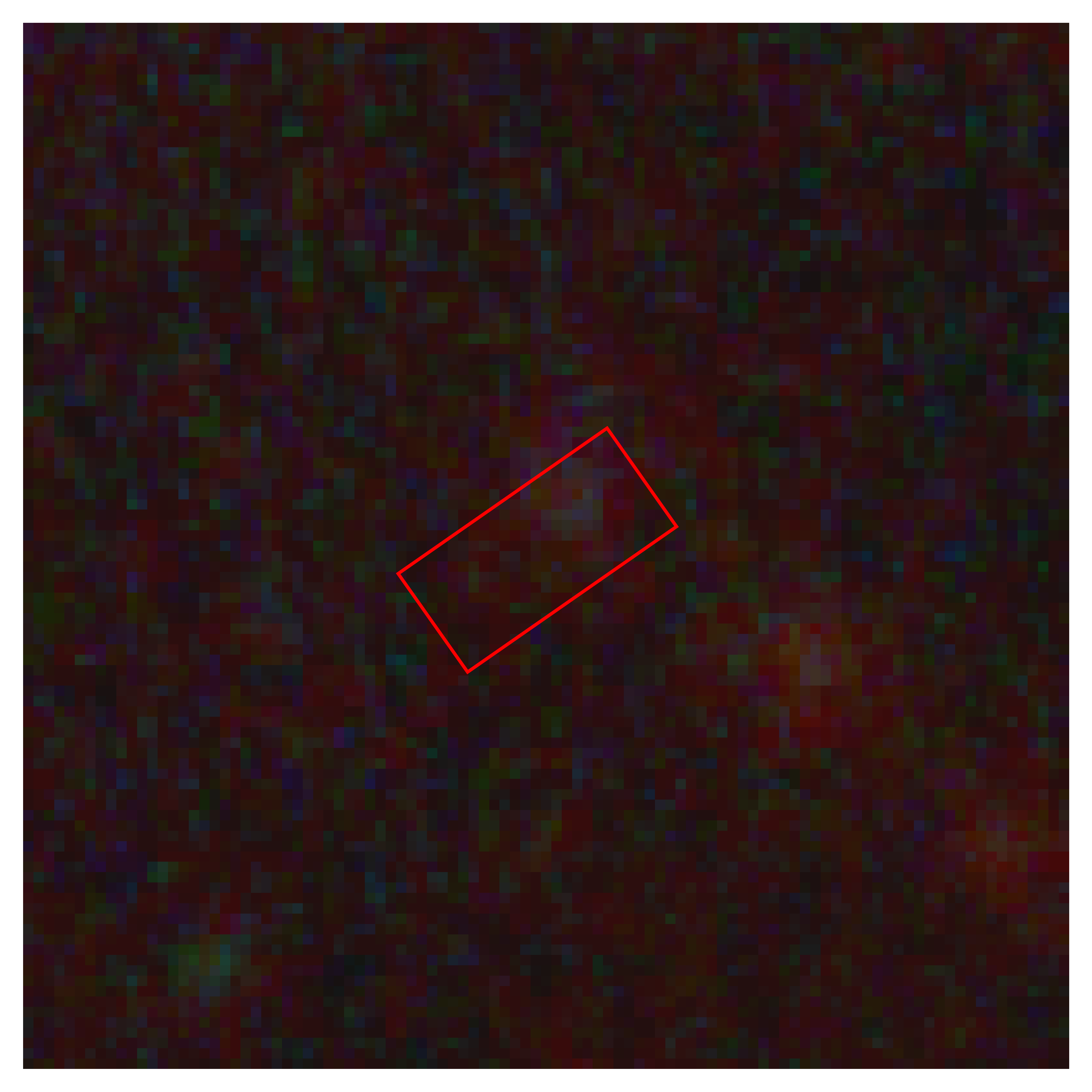}
\end{minipage}
\includegraphics[width=0.50\hsize]{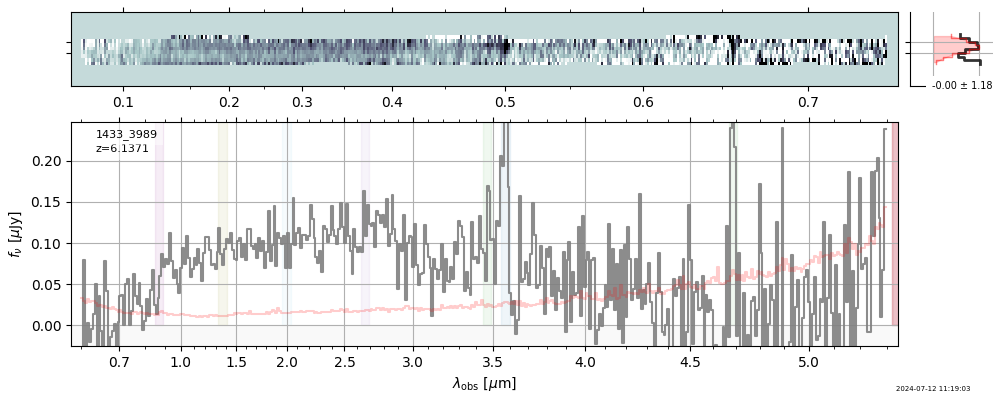}
\includegraphics[width=0.37\hsize, height=1.5in]{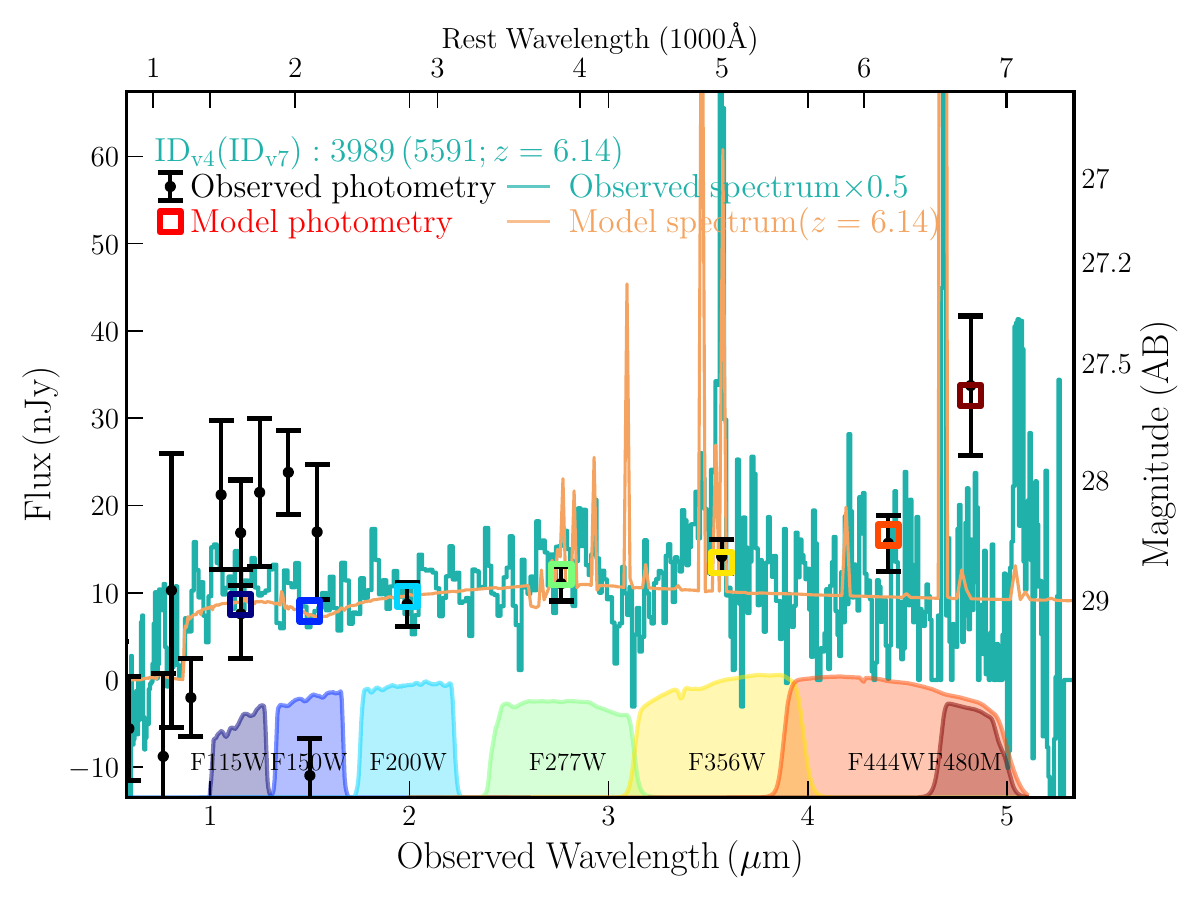}

\begin{minipage}[b]{0.12\hsize}
  \raggedright
  v4 ID 1715\\
  v7 ID 2121\\
  $z = 6.14$
  \includegraphics[width=\hsize]{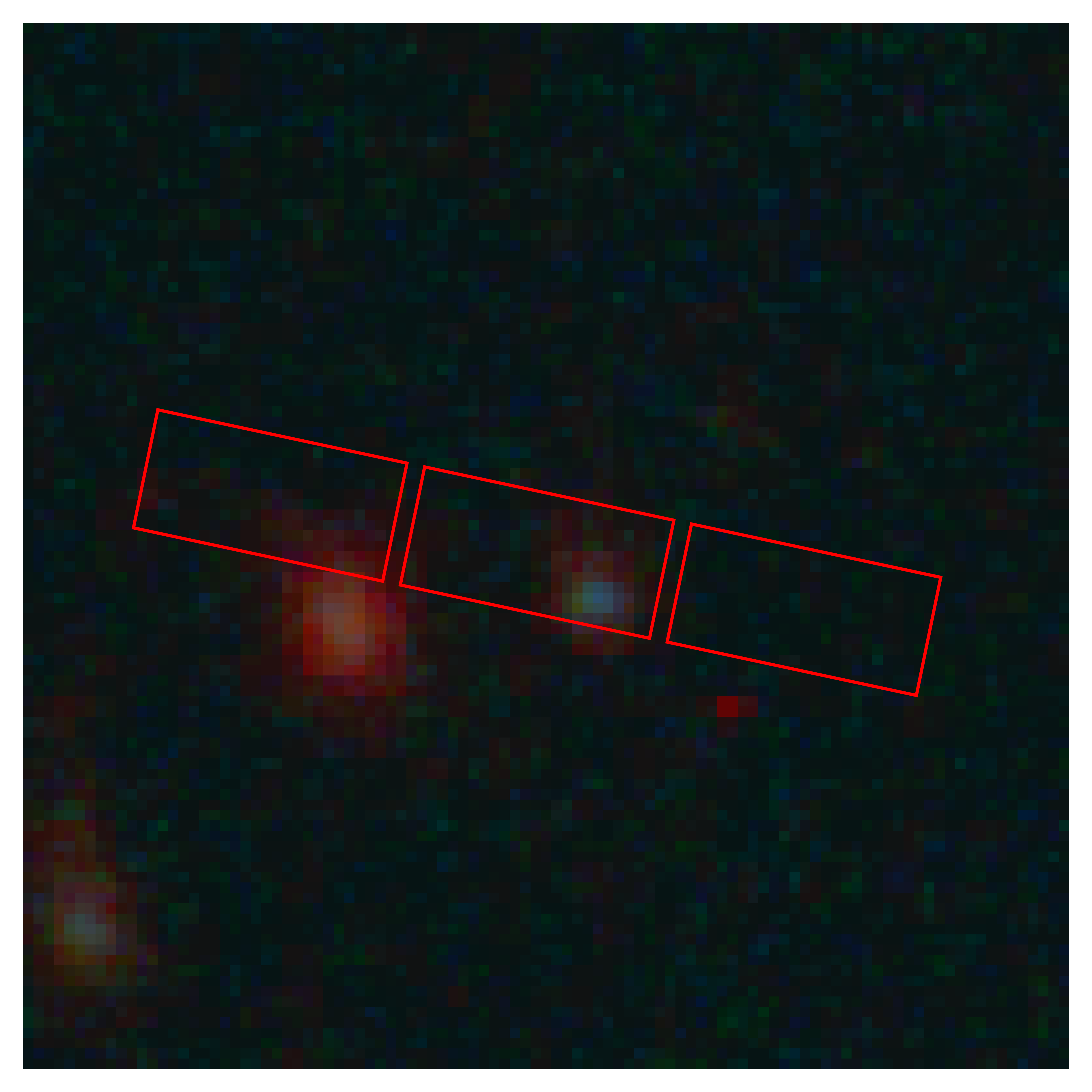}
\end{minipage}
\includegraphics[width=0.50\hsize]{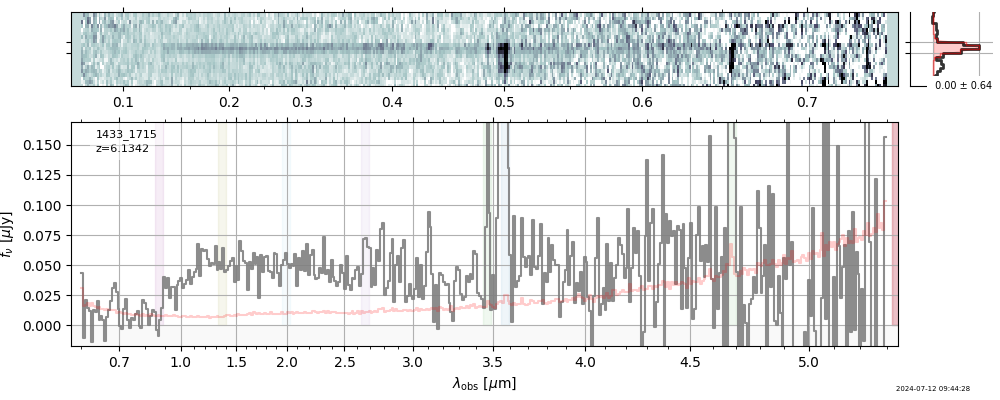}
\includegraphics[width=0.37\hsize, height=1.5in]{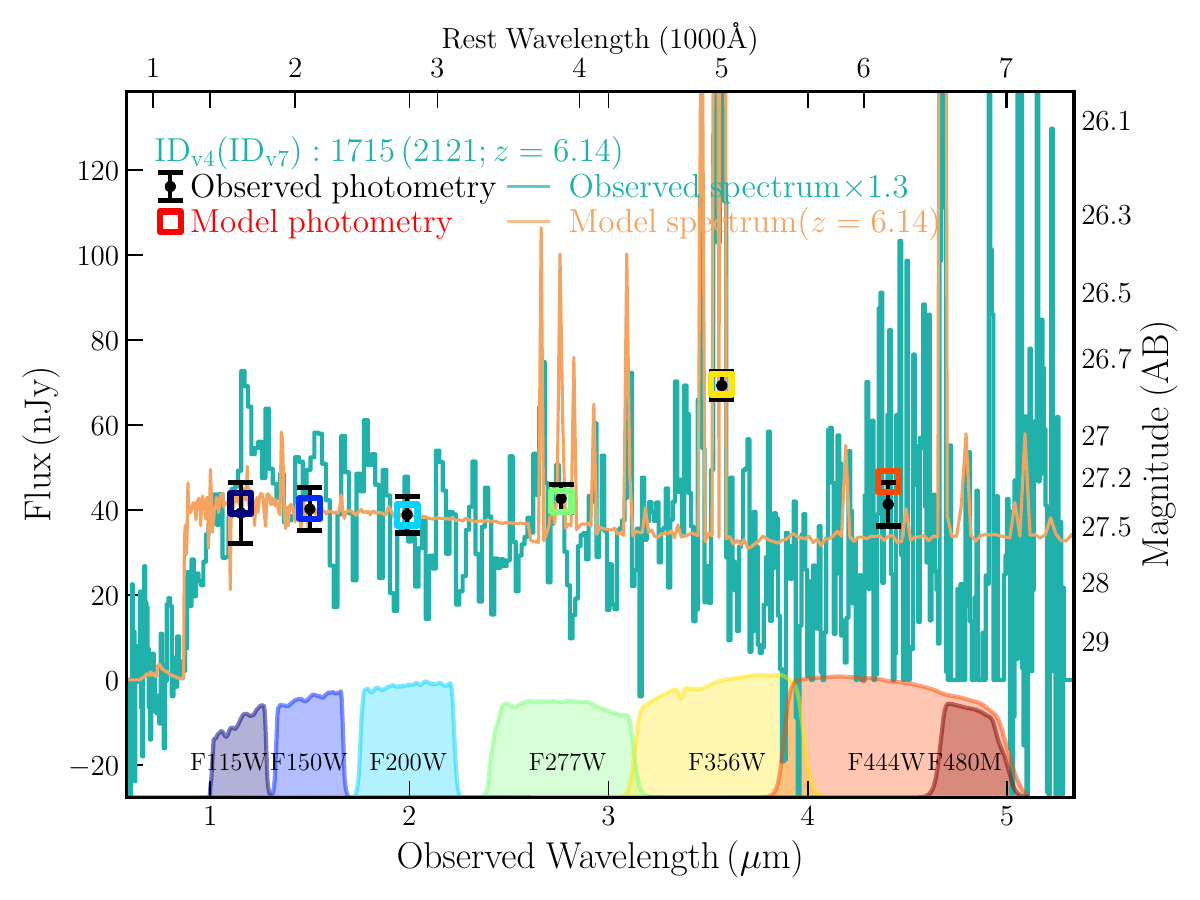}

\begin{minipage}[b]{0.12\hsize}
  \raggedright
  v4 ID 3308\\
  v7 ID 4533\\
  $z = 6.13$
  \includegraphics[width=\hsize]{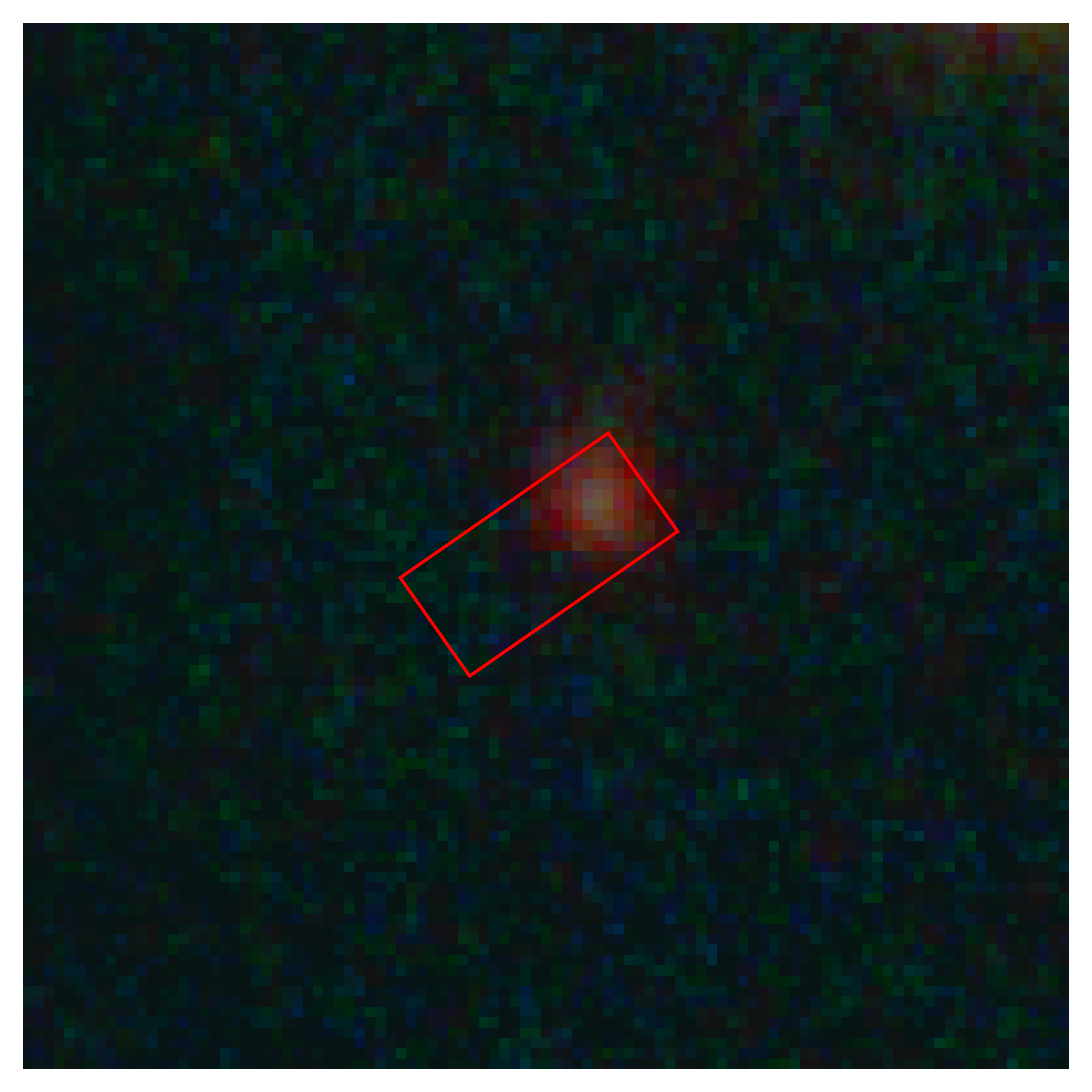}
\end{minipage}
\includegraphics[width=0.50\hsize]{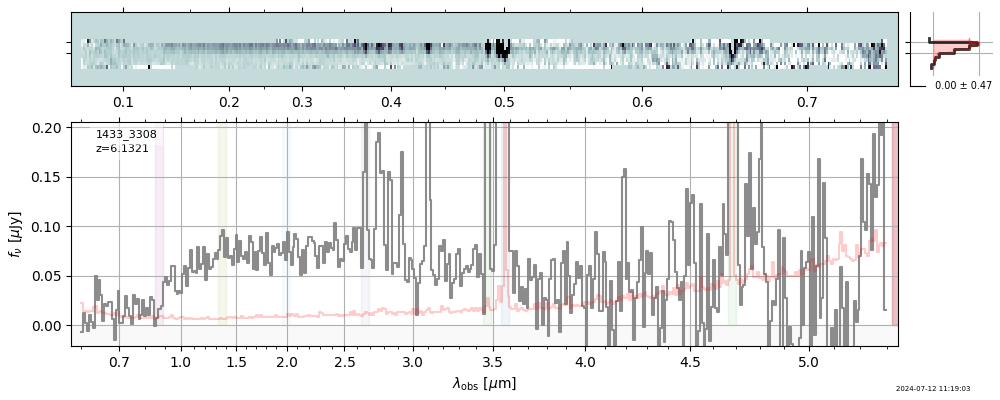}
\includegraphics[width=0.37\hsize, height=1.5in]{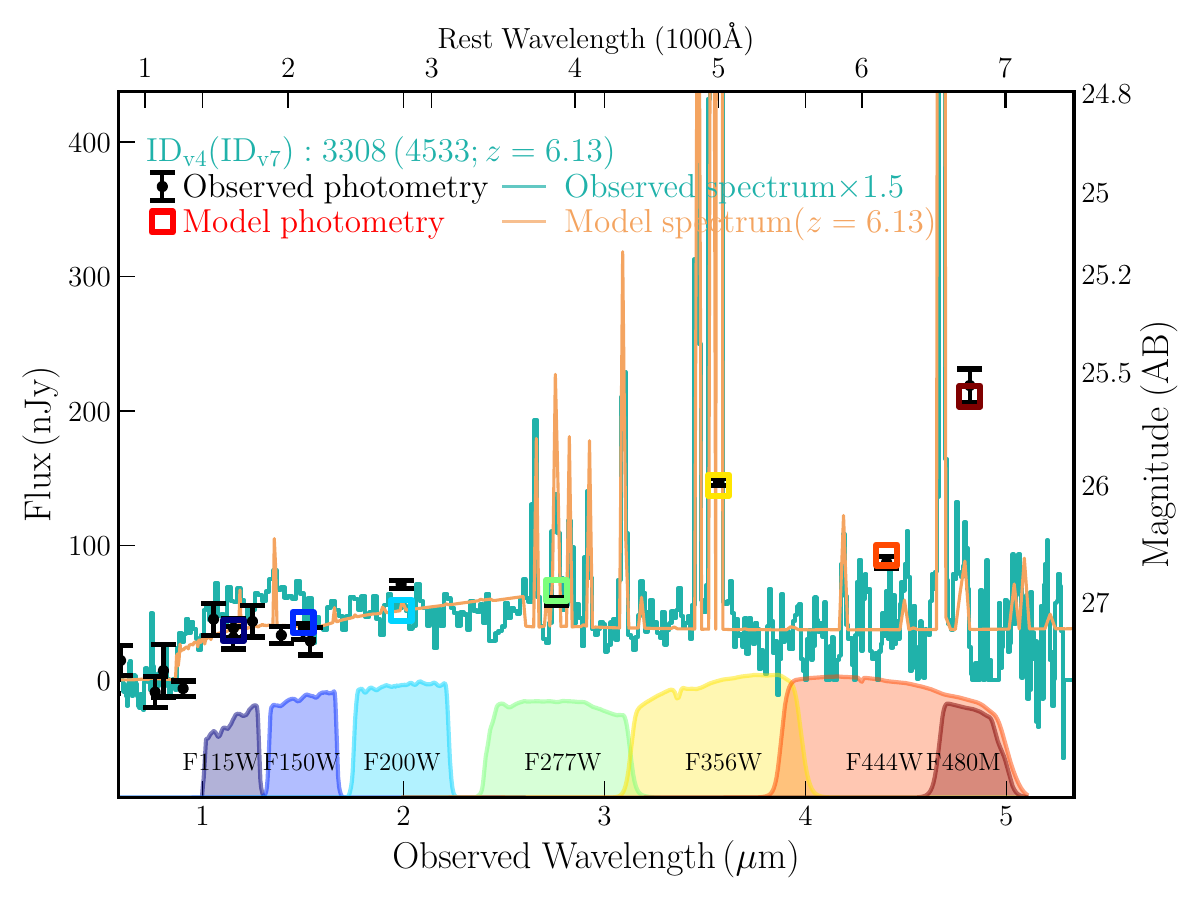}

\begin{minipage}[b]{0.12\hsize}
  \raggedright
  v4 ID 3208\\
  v7 ID 4411\\
  $z = 6.12$
  \includegraphics[width=\hsize]{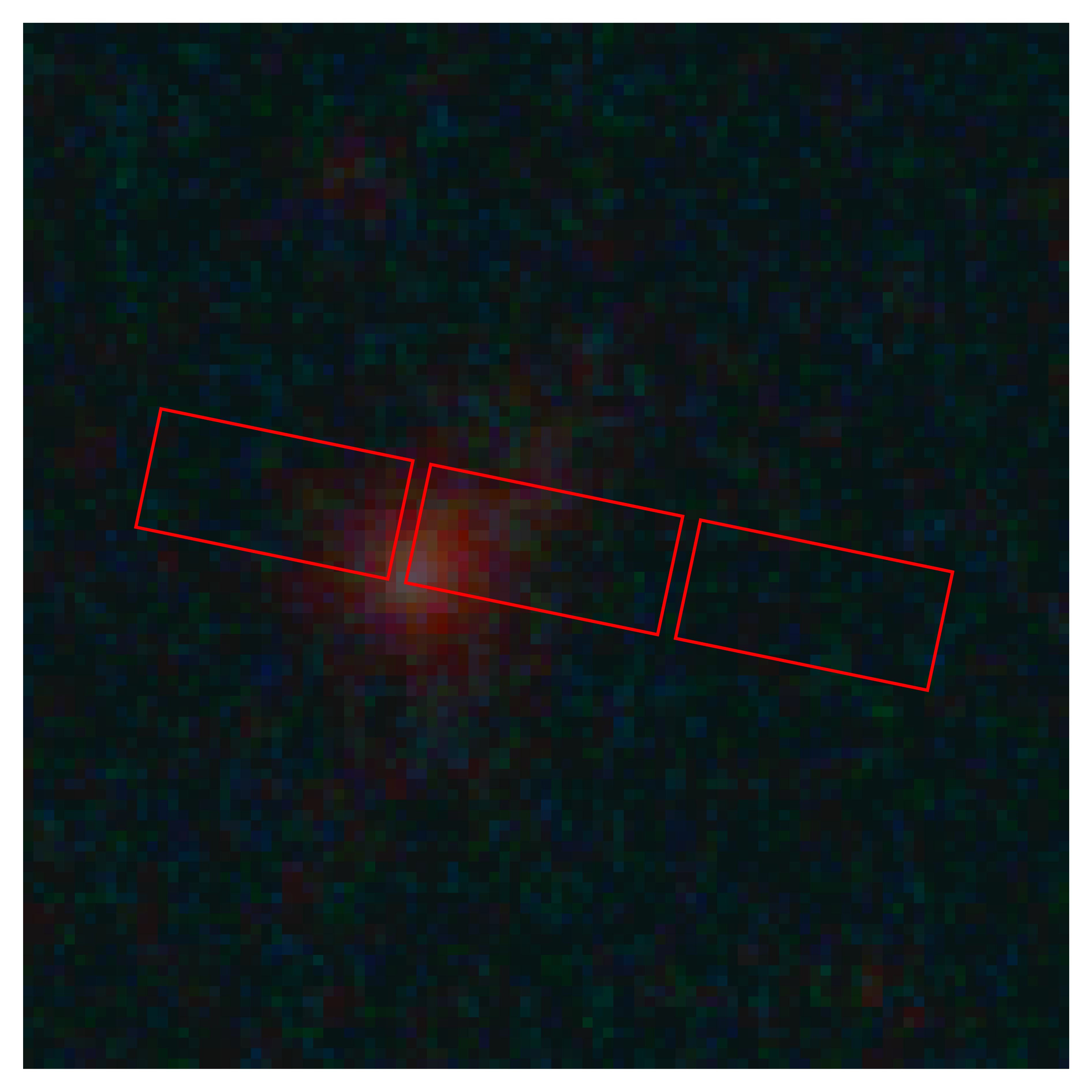}
\end{minipage}
\includegraphics[width=0.50\hsize]{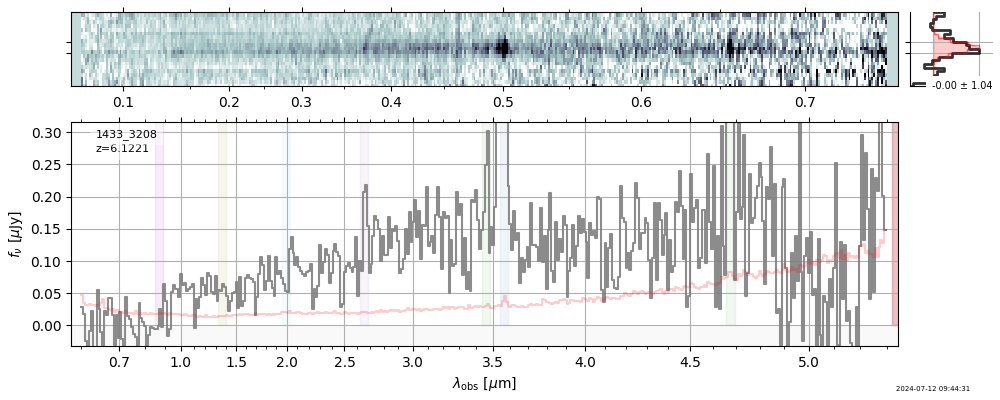}
\includegraphics[width=0.37\hsize, height=1.5in]{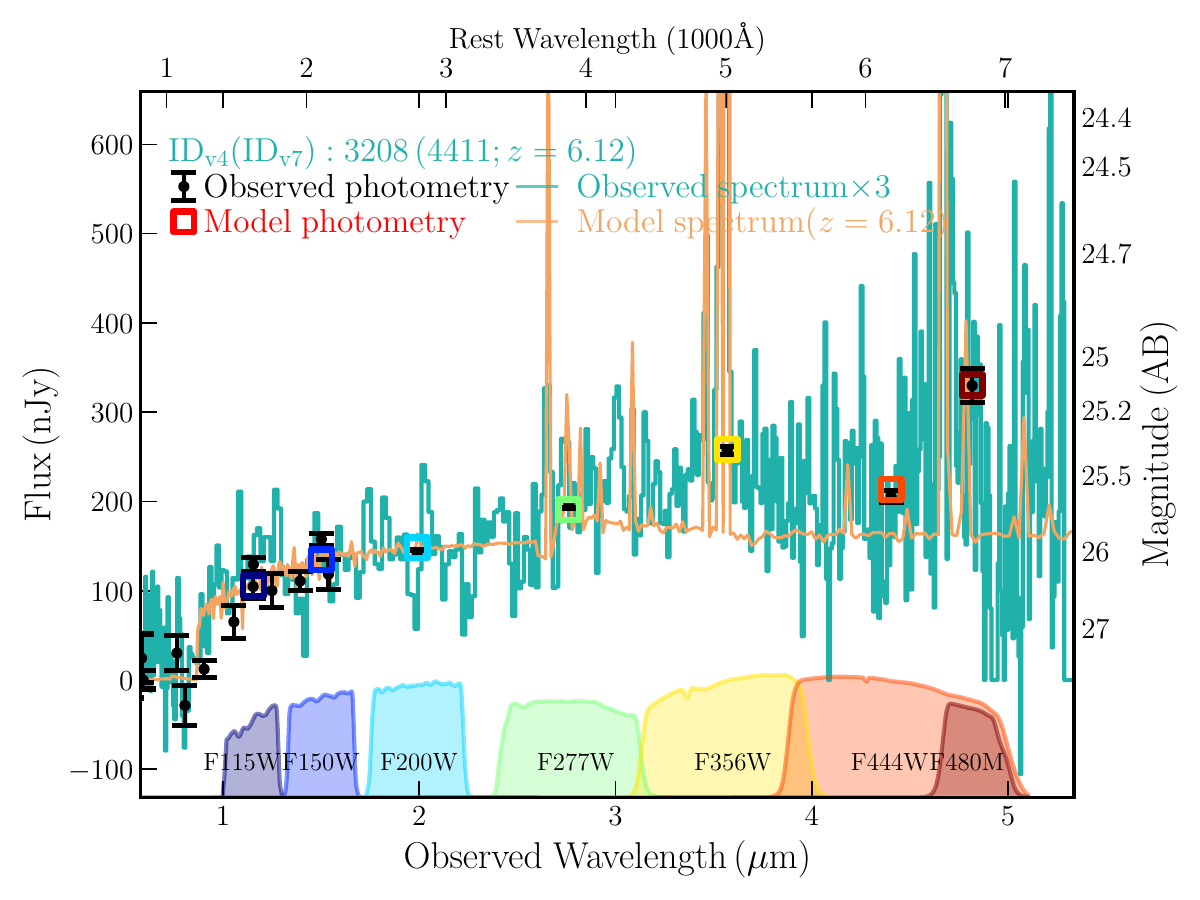}

\caption{
Four example members of the overdensity at $6.12 < z < 6.15$ identified with NIRSpec PRISM. For each galaxy we show \textbf{Left:} the NIRCam color image with slit orientation, \textbf{Middle:} the 2D and 1D PRISM spectra, and \textbf{Right:} the NIRCam photometry together with the SED model constrained by the spectrum. These sources, together with five additional galaxies in the same narrow redshift slice, indicate a prominent overdensity at $z \sim 6.1$ behind MACS0647.}
\label{fig:prism_z61}
\end{figure*}

\subsection{Spectral Energy Distribution (SED) Fitting} 
\label{sec:sedfitting}
We completed the SED fitting through the analysis codes \textsc{BAGPIPES} (Bayesian Analysis of Galaxies for Physical Inference and Parameter EStimation)~\citep{Carnall2018_BAGPIPES} and EAZY (Easy and Accurate Z-photo from Yale) ~\citep{Brammer2008_EAZY} with redshift as a free parameter.

Our initial SED fitting was performed with EAZY, which features a non-negative linear combination of templates from HST, Spitzer, and JWST photometry. We use the FSPS (Flexible Stellar Population Synthesis) templates~\citep{Conroy2010_SFPS} in the fitting. While these templates are widely used for high-redshift studies, they are calibrated primarily on low-redshift galaxies, and uncertainties in stellar population models at $z > 4$ can affect parameter estimates~\citep{Mobasher2015, Leja2019}. To mitigate these uncertainties, we performed follow-up analysis with \textsc{BAGPIPES}. 

\begin{figure*} [ht]
\begin{minipage}[b]{0.12\hsize}
  \raggedright
  v4 ID 3568\\
  v7 ID 4922\\
  $z = 9.25$
  \includegraphics[width=\hsize]{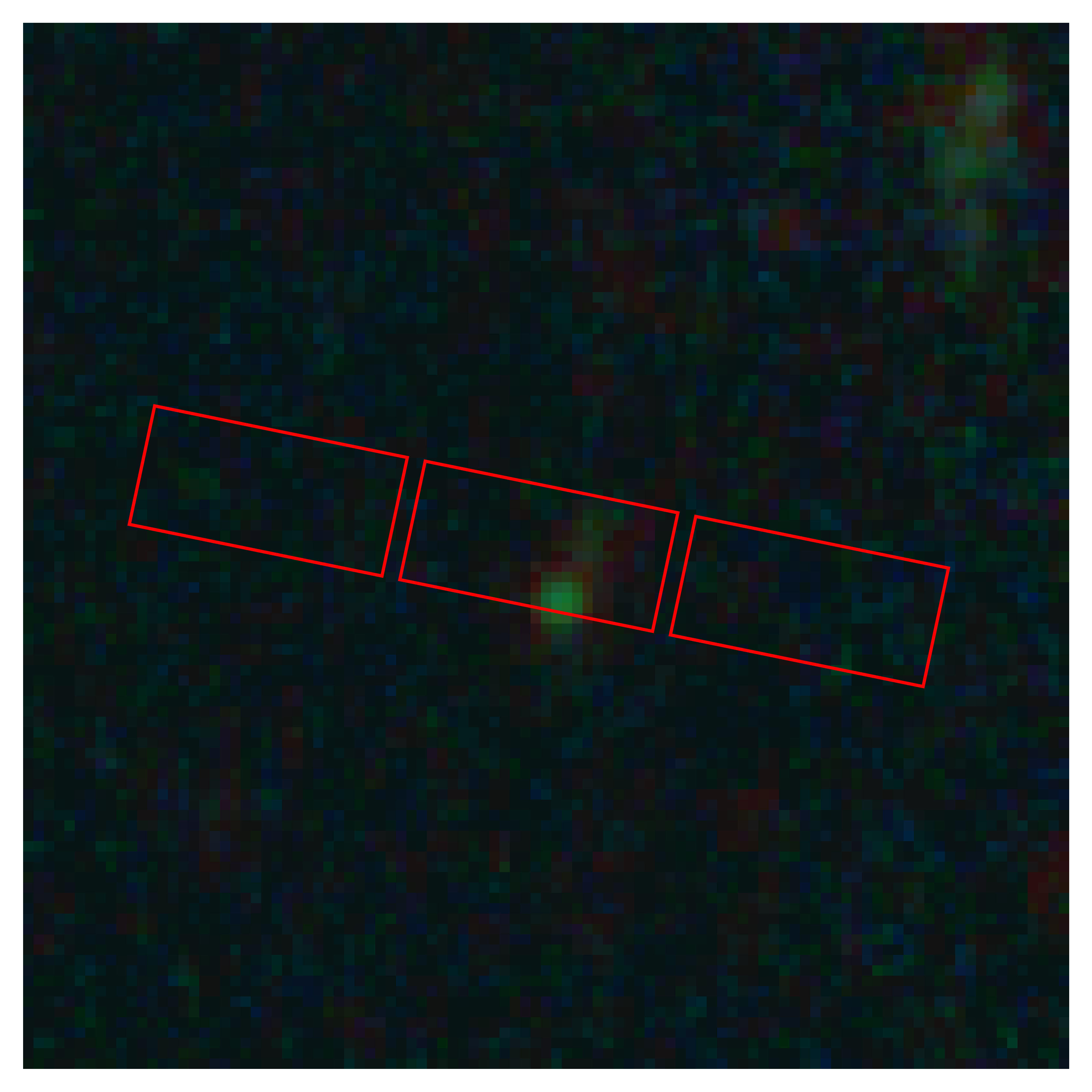}
\end{minipage}
\includegraphics[width=0.50\hsize]{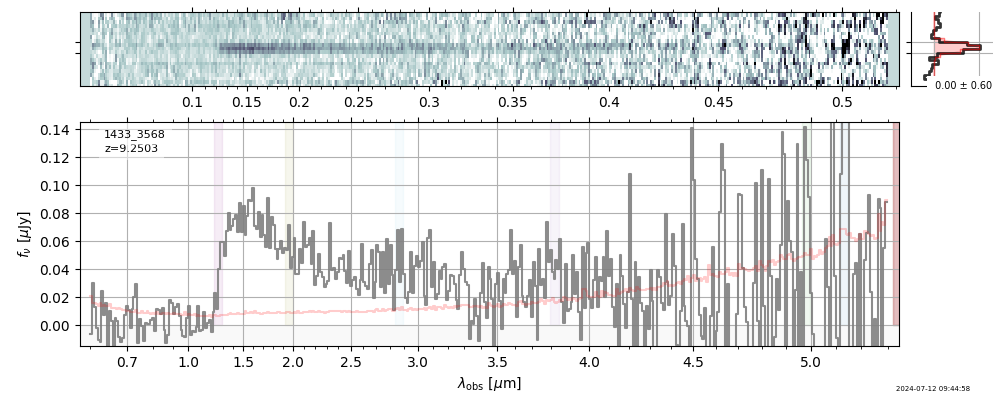}
\includegraphics[width=0.37\hsize, height=1.5in]{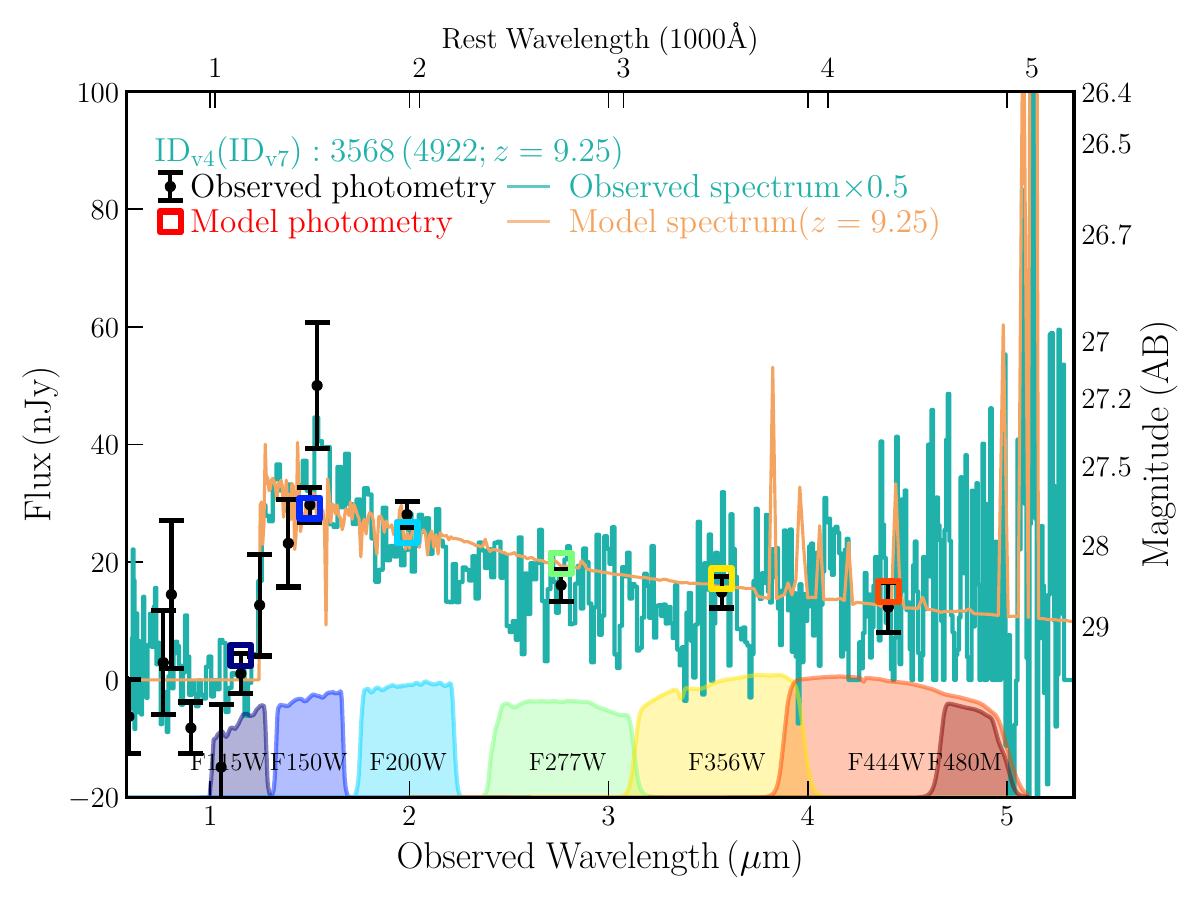}

\caption{.
Galaxy v4 3568 (v7 4922) at $z=9.25$ with NIRSpec PRISM, reported previously by \citep{Yanagisawa2024} and dubbed EBG-1 (Extremely Blue Galaxy). The source is modestly magnified by the foreground cluster, with $\mu = 2.72^{+0.09}_{-0.09}$. \textbf{Left:} NIRCam color images with slits overlaid. \textbf{Middle:} 2D and 1D spectra from DJA v3. \textbf{Right:} NIRCam photometry and SED fit with NIRSpec spectrum scaled to match.
}
\label{fig:prism_z79}
\end{figure*}

\textsc{BAGPIPES} is a Bayesian spectral-fitting code that uses the MultiNest nested sampling algorithm~\citep{Feroz2008_MultiNest, Feroz2009_MultiNest, Feroz2013_MultiNest} implemented through the PyMultiNest interface~\citep{Buchner2014_PyMultiNest} to estimate galaxy properties from photometric and spectroscopic data. The stellar population synthesis modeling is based on the 2016 updated BC03 models~\citep{Bruzual2003}, built on a Kroupa~\citep{Kroupa2002} initial mass function (IMF). 

For our SED fitting, we used two star formation histories, a constant star formation rate and a delayed exponentially declining SFH, where the star formation rate (SFR) is of the form SFR$(t)\propto t \exp{(-t/\tau)}$. The constant SFH model often yields systematically younger ages, whereas the delayed-$\tau$ model is generally considered better suited for high-redshift candidates due to its ability to capture rising star formation histories expected in early galaxies \citep[e.g.,][]{Papovich2011, Pacifici2013, Carnall2019}. However, we include the constant SFH for straightforward comparison with previous studies in the literature. We assume a Small Magellanic Cloud (SMC) dust law (Salim et al. 2018), with dust extinction ranging between $A_V$ = $0 - 3$. The age varied between 1 Myr and the age of the universe, and the permitted metallicity ranges were $\log Z/\Zsun = 0.005 - 5$. 

\subsection{Spectral Fitting} 
\label{sec:spectralfitting}
We measured emission line fluxes with \msaexp, which jointly models continuum plus Gaussian emission lines for each spectrum. The resulting line flux measurements from the PRISM data and the G395H observations are reported in Tables~\ref{tab:reversed_emission_lines} and \ref{tab:spec_g395h}, respectively.

\section{Results and Discussion} 
\label{sec:results&discussion}
Based on SED fitting to the NIRCam photometry, we present estimates of physical parameters in Table \ref{tab:catalog}. We then turn to the spectroscopy to measure properties in more detail.

\subsection{Spectroscopic Results}
In Table~\ref{tab:emission_lines}, we present the measured emission line fluxes for 9 galaxies observed with NIRSpec PRISM. A total of 20 emission lines were detected across these galaxies, with each galaxy exhibiting between 4 and 13 lines in their spectra. All galaxies in the sample have the [OIII] $\lambda$5007 line, an important tracer of star-formation. For four of the galaxies, we also report measurements of metallicity and the ionization parameter, providing additional constraints on their chemical composition and ionization states. 

The physical properties inferred from both photometry and spectroscopy indicate a population of relatively low-mass, low-metallicity galaxies. We measure metallicities of $\sim 10$--$40\%$ $Z_\odot$ and stellar masses between $\sim 10^8$ and $10^9\,M_\odot$. These values are broadly consistent with other recent studies of galaxies at $z > 6$ using JWST and ground-based spectroscopy~\citep{CurtisLake2023, Sanders2023, Cameron2024, Hainline_2024, Heintz_2025}. The low metallicities reflect the limited time for chemical enrichment in the early universe, while the stellar masses and star formation rates indicate that our sample includes both relatively young, intensely star-forming systems and more quiescent objects. The SFRs derived from our SED fits also span a wide range, suggesting a diversity of evolutionary stages among the high-redshift population.

It is important to note that our spectroscopic sample is subject to selection effects. The NIRSpec targets were chosen based on photometric redshift and brightness criteria, which preferentially select galaxies with strong rest-optical emission lines or prominent breaks. This selection can bias the sample toward actively star-forming or less dusty galaxies, and the physical properties we derive may not be fully representative of the underlying population of all $z>6$ galaxies in the field~\citep{Tang2024, Rhoads2023}. Additionally, uncertainties in lensing magnification factors, particularly for multiply imaged sources, can affect the intrinsic luminosities and derived physical parameters~\citep{Welch2022, Meena2023}.

\subsection{Extremely Blue Galaxy EBG-1 at $z = 9.25$}

Our highest-redshift galaxy, v4~3568 (v7~4922), lies at $z = 9.25$ and was previously identified as EBG-1 by \citet{Yanagisawa2024}. This ``extremely blue galaxy'' ($\beta = -3.0$) is inferred to have a high ionizing photon escape fraction, $f_{\rm esc}^{\rm ion} > 0.5$, suggesting vigorous, low-metallicity star formation. Figure~\ref{fig:prism_z79} presents the NIRCam imaging, NIRSpec PRISM spectroscopy, and SED fit for EBG-1.

In our analysis, the JWST spectrum exhibits a turnover in $F_\nu$ that may arise from either damped Ly$\alpha$ absorption \citep{Heintz2024} or two-photon continuum emission associated with a young, metal-poor stellar population and a top-heavy IMF \citep{Cameron2024}. Both scenarios point to an early, rapidly evolving system.

EBG-1 is only modestly magnified in our lensing model, with $\mu \approx 2.7$ (Table~\ref{tab:catalog2}), indicating that its extreme UV slope and high inferred escape fraction are intrinsic properties rather than artifacts of strong lensing. This makes EBG-1 one of the most compelling $z>9$ galaxies in the field, and a prime target for future spectroscopic and radiative-transfer studies.

\subsection{High-redshift $z \sim 7$ galaxies}
\begin{figure*} 
\centering
\includegraphics[width=0.90\textwidth]{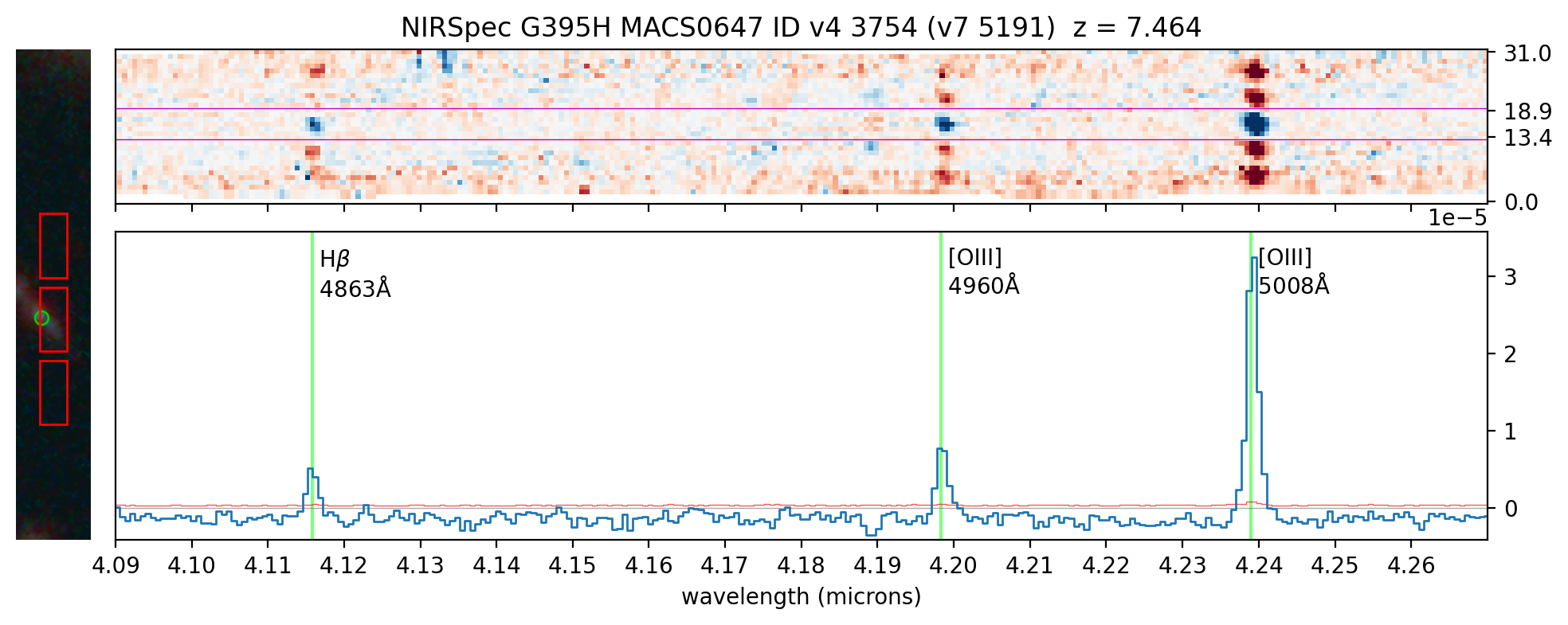}
\caption{
NIRSpec G395H high-resolution spectrum of galaxy v4 3754 (v7 5191) at $z = 7.464$, showing resolved \Hb, \OIII, and \Ha. The full 2.9--5.0\,$\mu$m range also covers \OIIw, which is not detected, consistent with the galaxy's high excitation.
}
\label{fig:3754_g395h}
\end{figure*}

We confirm two galaxies at $z \sim 7$. v4 3754 (v7 5191) at $z = 7.464$ features many emission lines in the PRISM data. \OIII\ and \Hb\ are also detected with G395H. v4 5094 (v7 7282) is confirmed at $z = 6.997$ with G395H detections of \OIII, \NeIII, and \OII. We include the G395H spectra of v4 3754 (v7 5191) within the Appendix~A (see Figure~\ref{fig:3754_g395h}).

\subsection{Overdensity at $z = 6.1$}
\begin{figure*} [htbp]
\includegraphics[height=4.5cm,keepaspectratio]{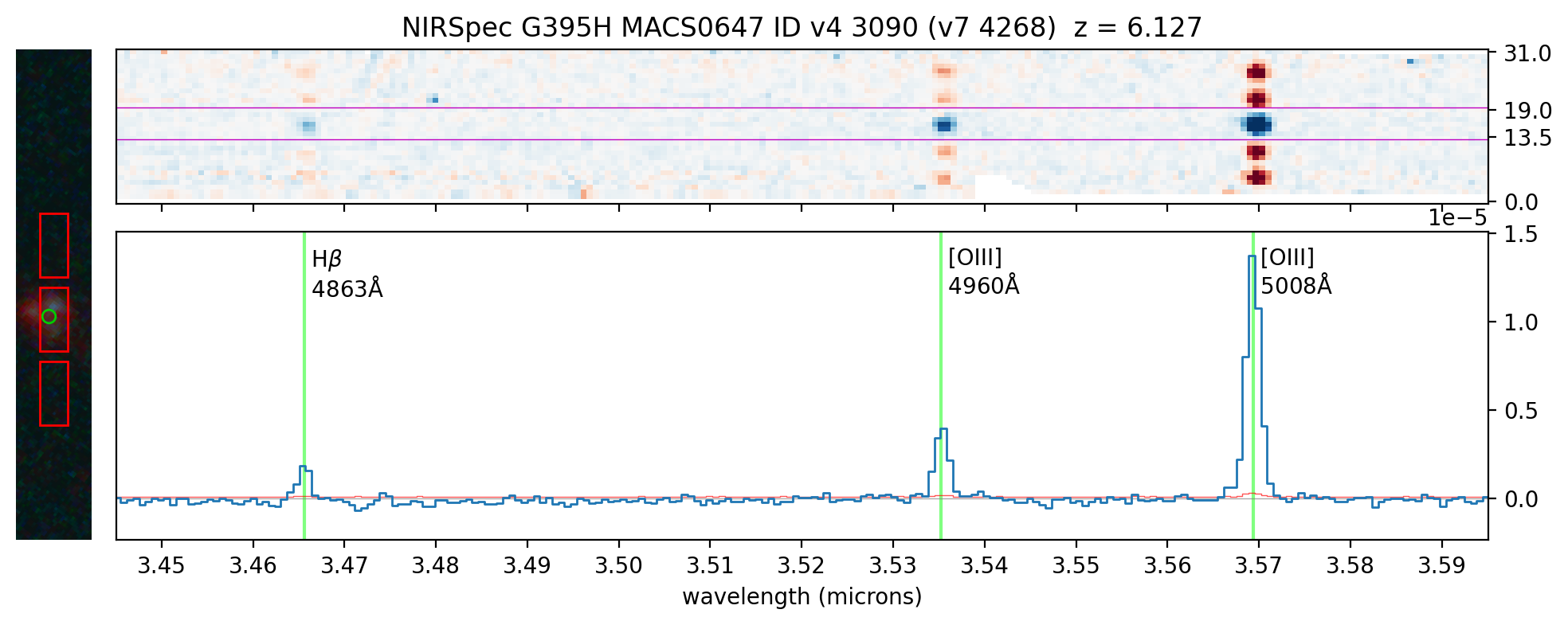}
\includegraphics[height=4.5cm,keepaspectratio]{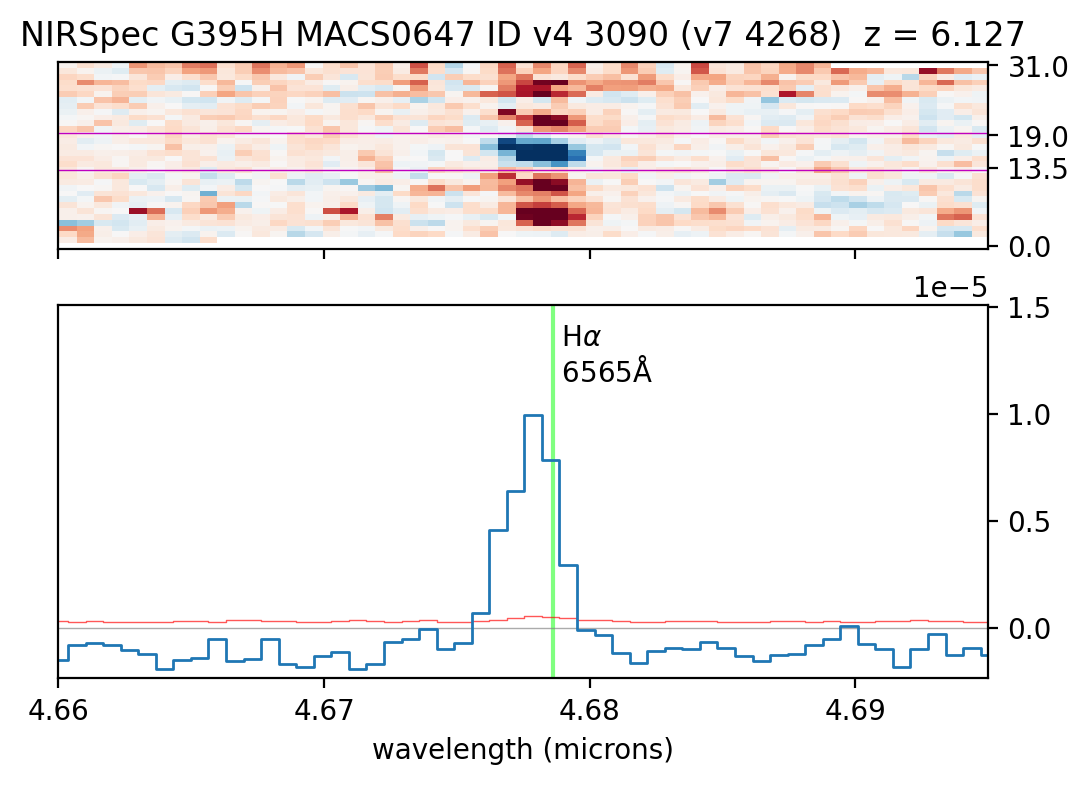}

\includegraphics[height=4.5cm,keepaspectratio]{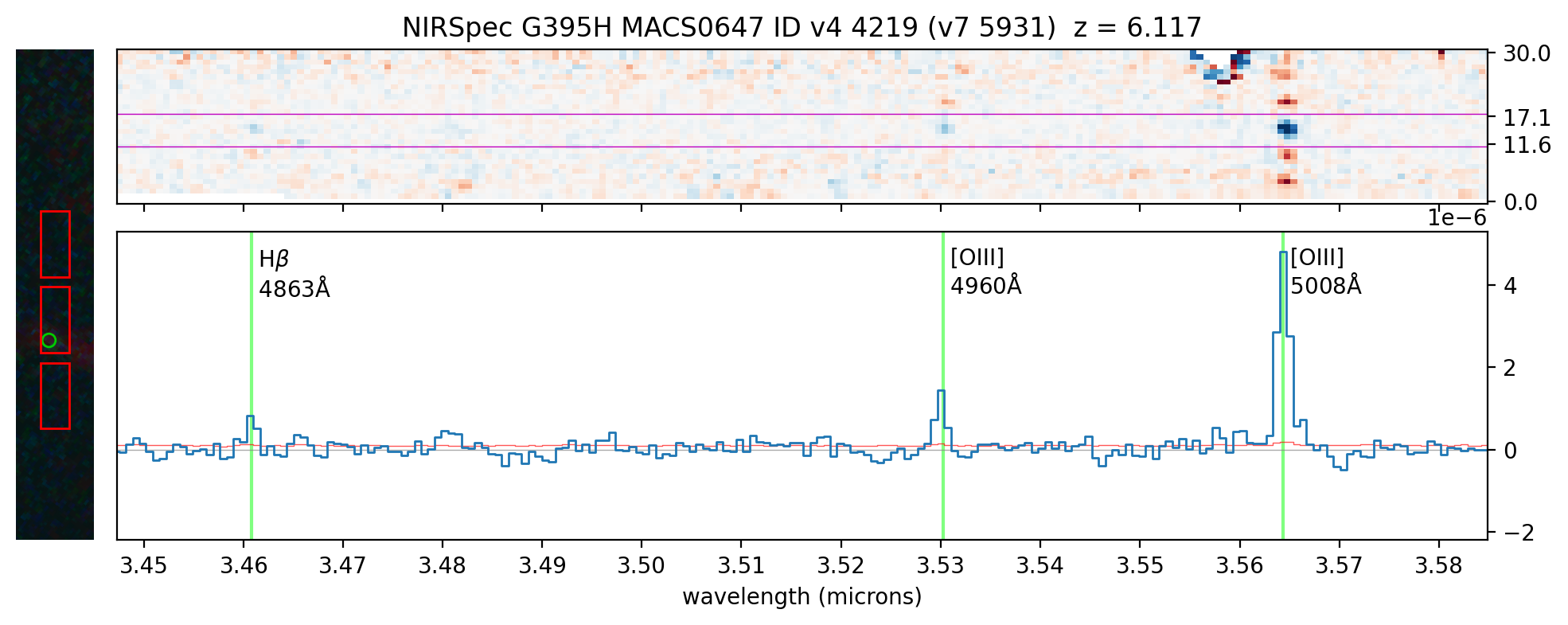}
\includegraphics[height=4.5cm,keepaspectratio]{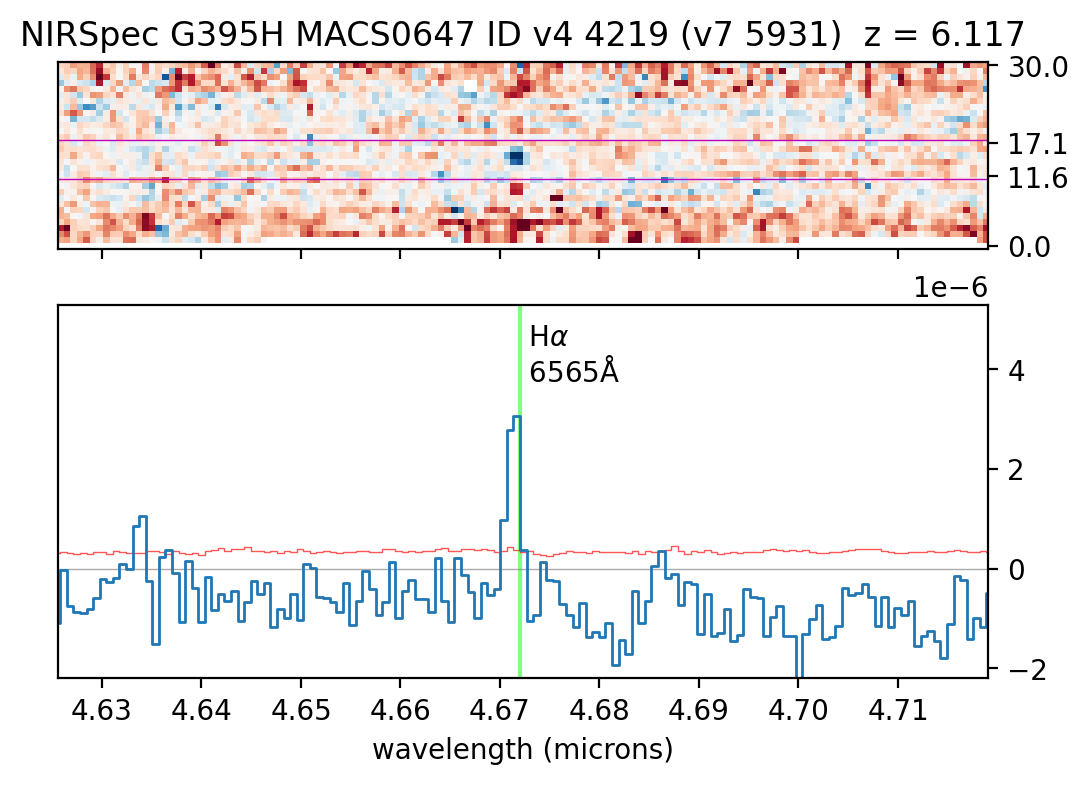}

\includegraphics[height=4.5cm,keepaspectratio]{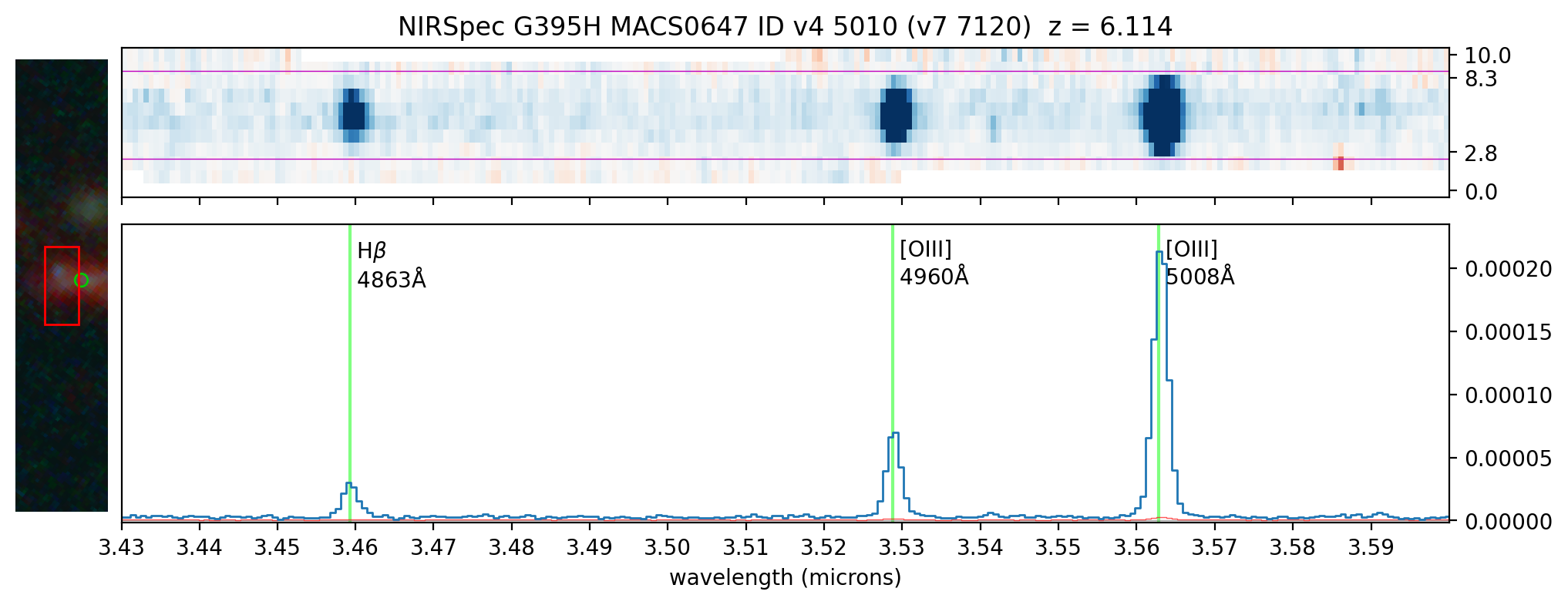}
\includegraphics[height=4.5cm,keepaspectratio]{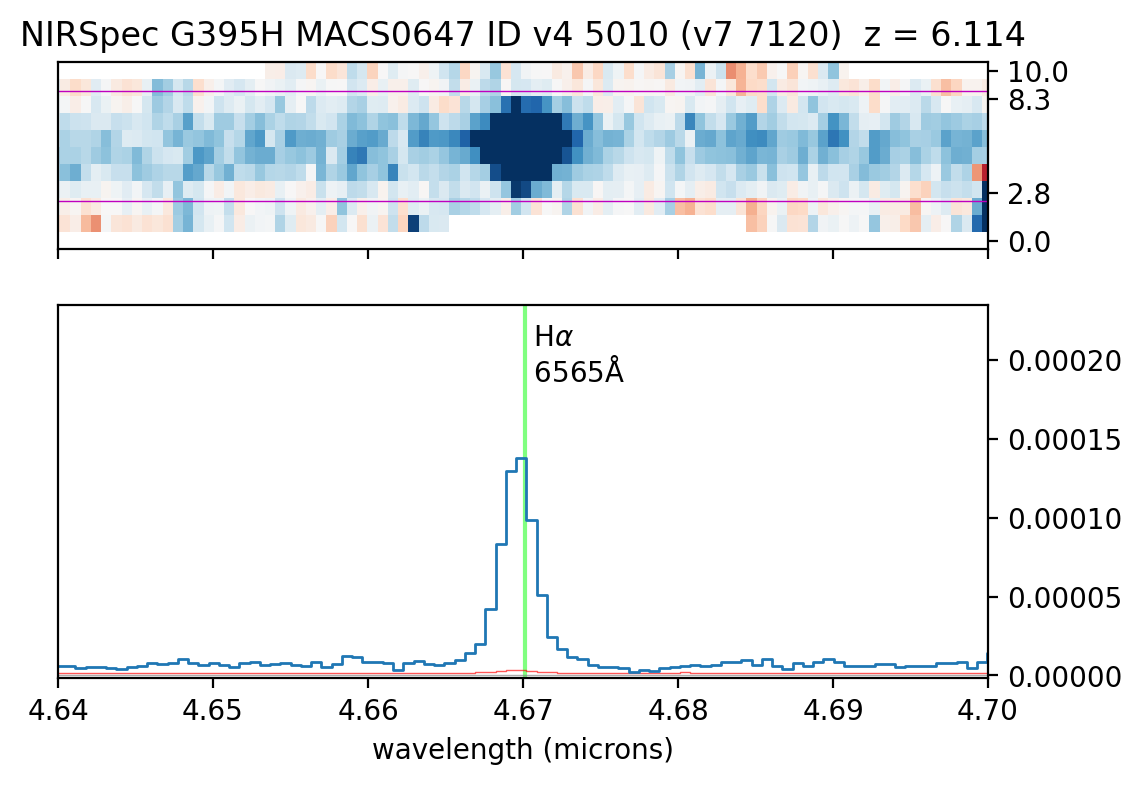}

\includegraphics[height=4.5cm,keepaspectratio]{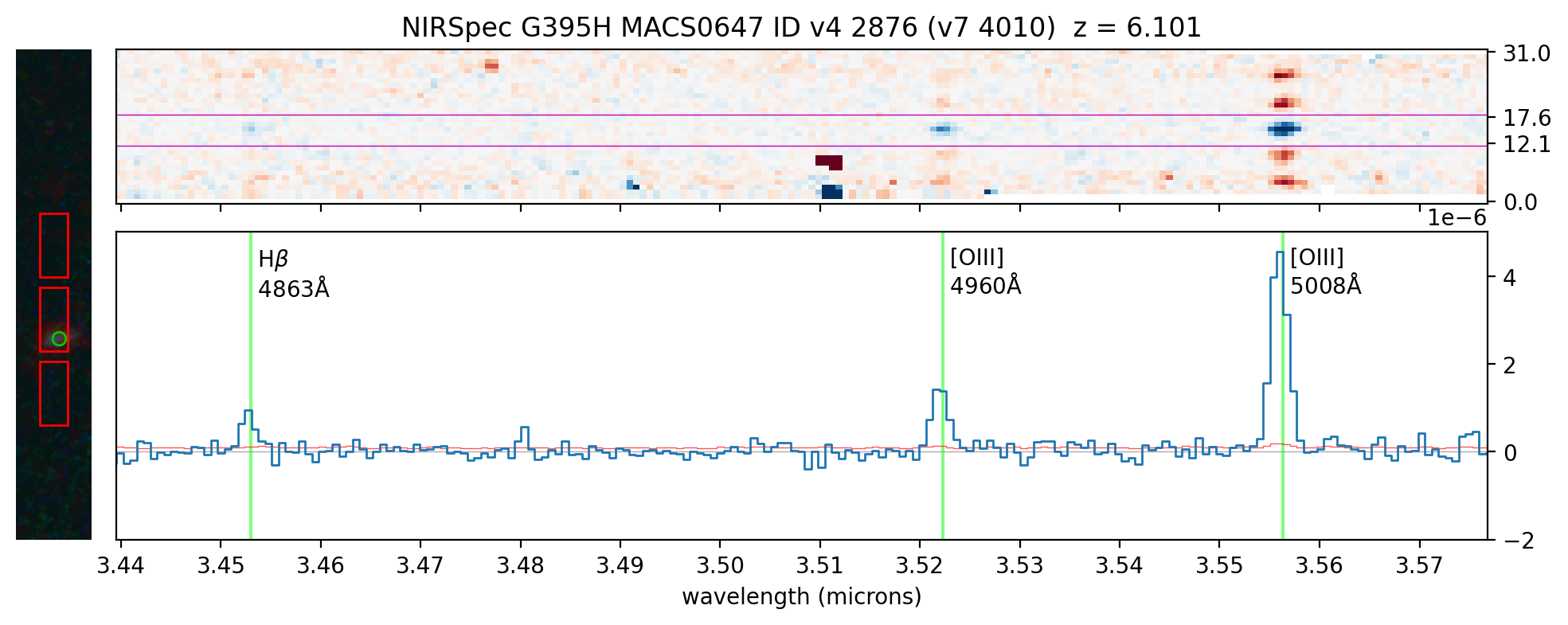}
\includegraphics[height=4.5cm,keepaspectratio]{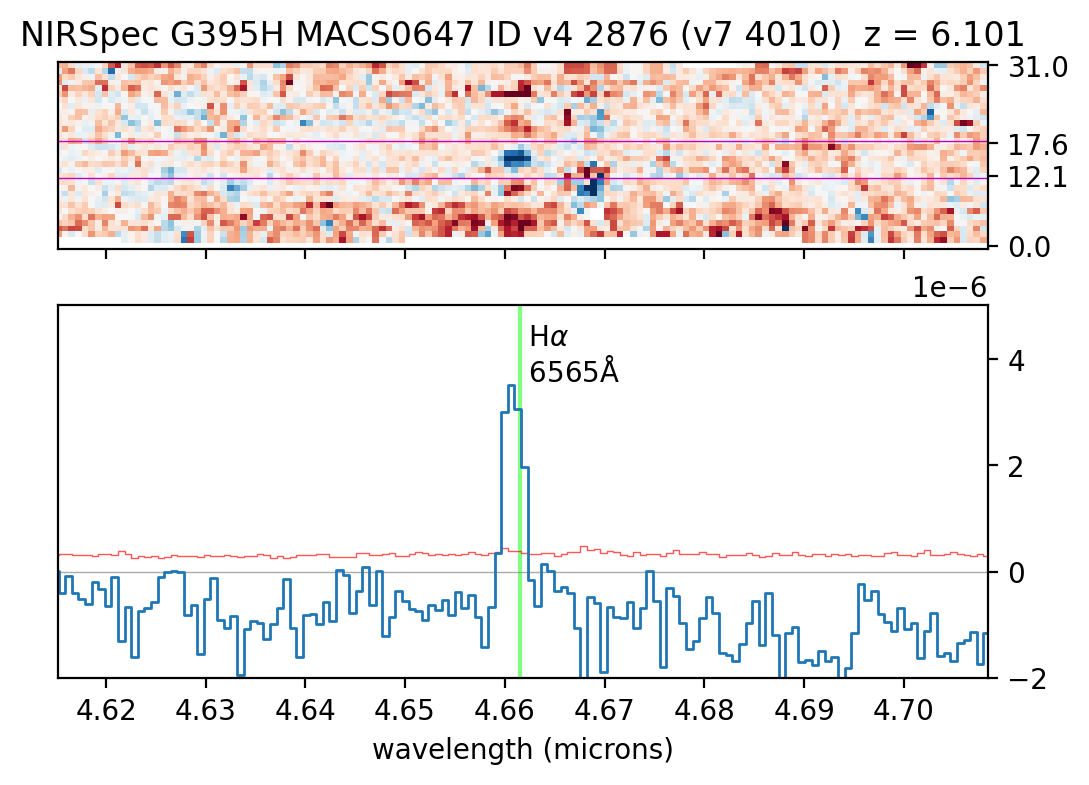}

\caption{
NIRSpec G395H high-resolution spectra of 4 galaxies that are part of our observed overdensity at $6.10 < z < 6.13$ showing \Hb, \OIII, and \Ha.
Note that v4 3090 (v7 4268) consists of two components separated by $\sim$1 kpc and $\sim$90 km/s, suggesting a dynamical mass $\sim$$10^8 M_\odot$.
}
\label{fig:g395h_z61}
\end{figure*}

We detect an overdensity of galaxies at $z = 6.1$. We confirm 9 galaxies spectroscopically with $6.11 < z < 6.15$ in our PRISM and G395H observations (Figures \ref{fig:prism_z61} and \ref{fig:g395h_z61}). This confirms the significant overdensity of $z \sim 6$ galaxies reported previously in \HST\ observations of MACS0647 that was much higher than found behind other galaxy clusters~\citep{Bradley2014}.

We note the photometric redshifts of these galaxies are slightly higher, $6.5 < z < 6.9$, as estimated previously~\citep{Bradley2014} based on \HST\ images and again now with the addition of \JWST\ images. Such slight overestimates have been reported previously~\citep{Heintz_2025, Hainline_2024, Fujimoto_2023} and are attributed primarily to damped Ly$\alpha$ wings not accounted for in photometric redshift fitting codes. Roughly 20 galaxy candidates within our catalog fall in this range for photometric redshifts. 

We further find that some of the $z \sim 6$ candidates presented in the \HST\ analysis turn out to be more likely at the cluster redshift $z = 0.584$. This is due to a photometric redshift degeneracy: the cluster elliptical red rising SED can sometimes appear more like a high-$z$ Lyman break in noisy \HST\ photometric data for galaxies with mag $\sim 27$ and fainter~\citep{Song_2012}. This degeneracy is resolved by higher SNR \JWST\ data extending to longer wavelengths.

Possible explanations for the observed overdensity include the presence of a forming protocluster or a large-scale structure along the line of sight~\citep{Jiang_2018, Franck_2016}. However, the lack of strong Ly$\alpha$ emission from multiple galaxies in the overdensity makes it difficult to conclusively identify it as a protocluster~\citep{Jiang_2018, Bacon_2021}. Alternatively, cosmic variance or projection effects could also contribute to the apparent concentration of galaxies at this redshift~\citep{Franck_2016, Shi_2019}. Further spectroscopic confirmation, particularly of Ly$\alpha$ emission and velocity dispersion, will be necessary to distinguish between these scenarios.
\begin{table*}[htbp]
\centering
\setlength{\tabcolsep}{8pt}
\caption{Physical properties of spec-$z$ confirmed galaxies derived from \textsc{bagpipes} NIRSpec PRISM line flux fitting. \textbf{Note:} Little Red Dots (LRDs) are marked with an asterisk ($^{*}$); their SED fits may be unreliable due to extreme colors or AGN contamination.} \label{tab:catalog_spec}
\begin{tabular}{lcccccc}
\hline
ID (v7) & ID (v4) & $z_{\rm spec}$ & Stellar Mass (log($M_{*}/M_\odot$)) & SFR ($M_{*}$/yr) & $A_{V}$ (mag) & Mass-weighted Age (Gyr) \\
\hline
4922 & 3568 & $9.25$ & $7.80^{+0.25}_{-0.25}$ & $2.85^{+0.58}_{-0.37}$ & $0.01^{+0.01}_{-0.02}$ & $0.01^{+0.01}_{-0.02}$ \\
5191 & 3754 & $7.47$ & $8.30^{+0.03}_{-0.04}$ & $20.19^{+1.47}_{-1.82}$ & $0.04^{+0.02}_{-0.03}$ & $0.00^{+0.00}_{-0.00}$ \\
5591 & 3989 & $6.14$ & $8.05^{+0.23}_{-0.14}$ & $0.66^{+0.17}_{-0.28}$ & $0.16^{+0.10}_{-0.14}$ & $0.15^{+0.09}_{-0.10}$ \\
2121 & 1715 & $6.14$ & $8.14^{+0.20}_{-0.24}$ & $3.26^{+0.70}_{-0.85}$ & $0.06^{+0.03}_{-0.06}$ & $0.03^{+0.02}_{-0.04}$ \\
4533 & 3308 & $6.13$ & $7.72^{+0.03}_{-0.04}$ & $5.32^{+0.32}_{-0.51}$ & $0.10^{+0.03}_{-0.04}$ & $0.00^{+0.00}_{-0.00}$ \\
4411 & 3208 & $6.12$ & $8.93^{+0.09}_{-0.09}$ & $15.89^{+2.06}_{-2.31}$ & $0.11^{+0.03}_{-0.04}$ & $0.04^{+0.01}_{-0.02}$ \\
2193$^{*}$ & 1742 & $5.81$ & $9.56^{+0.17}_{-0.07}$ & $205.30^{+65.46}_{-47.51}$ & $0.94^{+0.14}_{-0.06}$ & $0.01^{+0.00}_{-0.00}$ \\
1658 & 1395 & $3.72$ & $8.93^{+0.10}_{-0.05}$ & $1.95^{+0.33}_{-0.51}$ & $0.23^{+0.08}_{-0.11}$ & $0.38^{+0.13}_{-0.13}$ \\
4870 & 3533 & $3.23$ & $8.25^{+0.05}_{-0.05}$ & $2.94^{+0.13}_{-0.17}$ & $0.00^{+0.00}_{-0.01}$ & $0.05^{+0.01}_{-0.01}$ \\
\hline
\end{tabular}
\end{table*}
 
\subsection{Little Red Dots}
We identify several Little Red Dots (LRDs) in this field. One was reported previously at $z = 4.53$ \citep{Killi2024} based on the Cycle 1 NIRCam and NIRSpec PRISM data. This object (v4 1045) has broad \Ha, a signature of AGN accretion disk broad line regions, with an estimated black hole mass $8 \times 10^8 M_\odot$. The photometry and spectra exhibit two components, including high dust attenuation $A_V > 5$ at longer wavelength, as found in other LRDs. Our Cycle 2 NIRSpec G395H data confirms the redshift and other properties.

Additionally, 1 arcmin away, our G395H spectrum of a $z \sim 6.5$ candidate v4 461 (v7 556) reveals it to be an LRD candidate at $z = 4.77$ with faint detections of \OIII\ and \Ha\ (not shown); the detected lines are narrow, with no evidence for significant broadening.

Our Cycle 1 PRISM data also reveals a $z \sim 7.5$ candidate v4 1742 (v7 2193) to be an LRD candidate at $z = 5.81$. With the low resolution PRISM data, we do not detect significant broadening. The line is not clearly resolved with a upper limit width of $\sim$1000 km/s. But we do find that an additional broad component $\sim$4000 km/s significantly improves the fit. This may hint at the presence of an AGN.

\subsection{Lensing and Magnification}
MACS0647 is a well-studied strong-lensing cluster with several published mass models (e.g., \citealt{Zitrin2011, Coe2013, Nishida2025}). The depth and resolution of JWST reveal additional faint lensed sources as well as substructure within previously known images, enabling more precise lensing constraints.

Several of our high-redshift candidates correspond to systems identified in prior lensing analyses \citep{Welch2022, Meena2023}. We list the matching IDs in Tables~\ref{tab:catalog2}--\ref{tab:catalog2b}, providing a direct connection to existing mass-model constraints. Identifying these associations is essential for avoiding double counting in number-density estimates and for interpreting the intrinsic properties of these galaxies \citep{Barkana2000, Lotz2017}.

Magnification factors for all candidates were computed using the mass model presented in~\citep{Nishida2025} that uses 86 multiple images from 28 sources to construct the mass model with \textsc{glafic}~\citet{Oguri2010, Oguri2021}. Most galaxies lie outside the critical regions and therefore experience only mild magnification ($\mu \approx 1$--2) while a smaller subset located closer to the critical curves exhibit substantially higher values. Among the spectroscopically confirmed sources, the most magnified is ID~5931 ($z_{\rm spec}=6.12$) with $\mu \approx 7.4$. The highest magnification in the full sample is ID~5595, which reaches $\mu \approx 11$, consistent with previous analyses of MACS0647.

These magnification estimates are used when reporting intrinsic luminosities and for comparing to luminosity functions. While the majority of our sample is only weakly affected by lensing, the most highly magnified systems provide valuable leverage for probing intrinsically faint galaxies and represent the dominant sources of model-dependent systematic uncertainty in our analysis \citep{Meneghetti:2016hcr, Cerny2018}.

\subsection{Low Redshift Interlopers}
\begin{figure*} [!htp]

\begin{minipage}[b]{0.12\hsize}
  \raggedright
  v4 ID 1395\\
  photo-$z$ $\sim$ 12.5\\
  v7 ID 1658\\
  $z = 3.72$
  \includegraphics[width=\hsize]{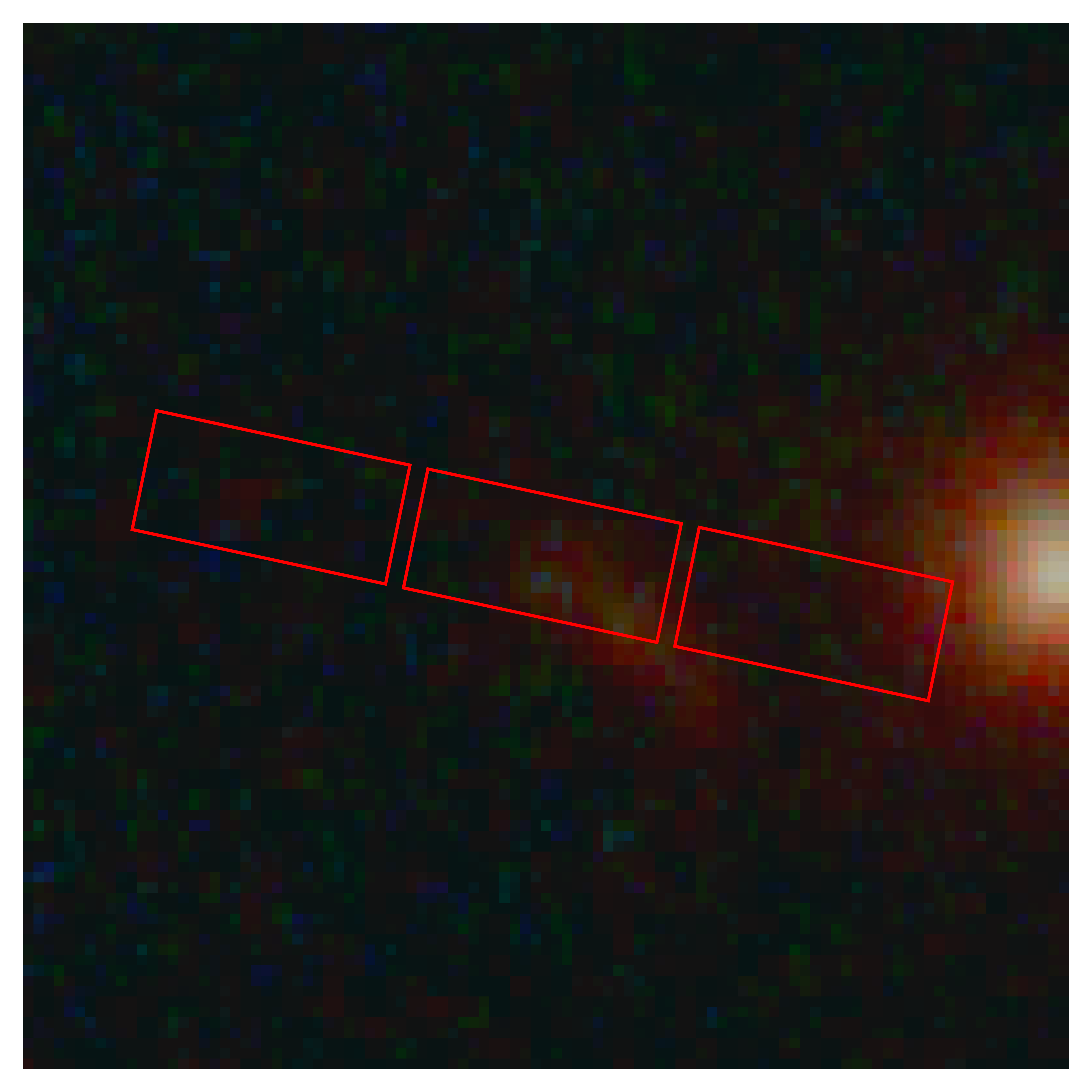}
\end{minipage}
\includegraphics[width=0.50\hsize]{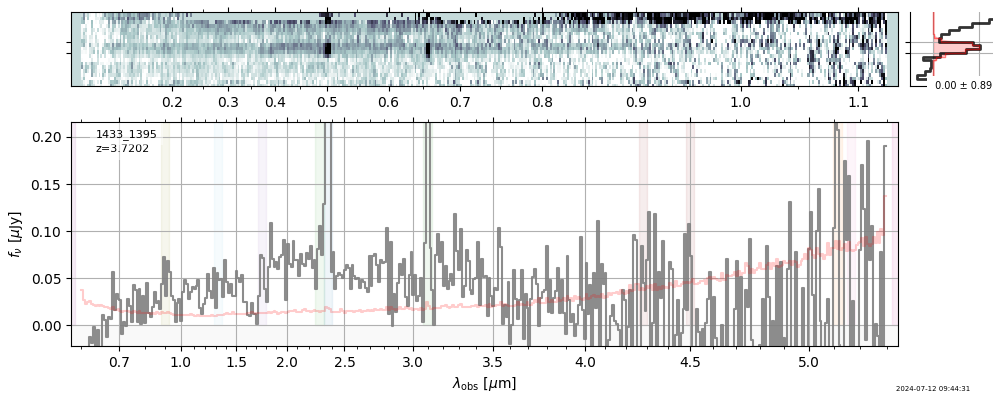}
\includegraphics[width=0.37\hsize, height=1.5in]{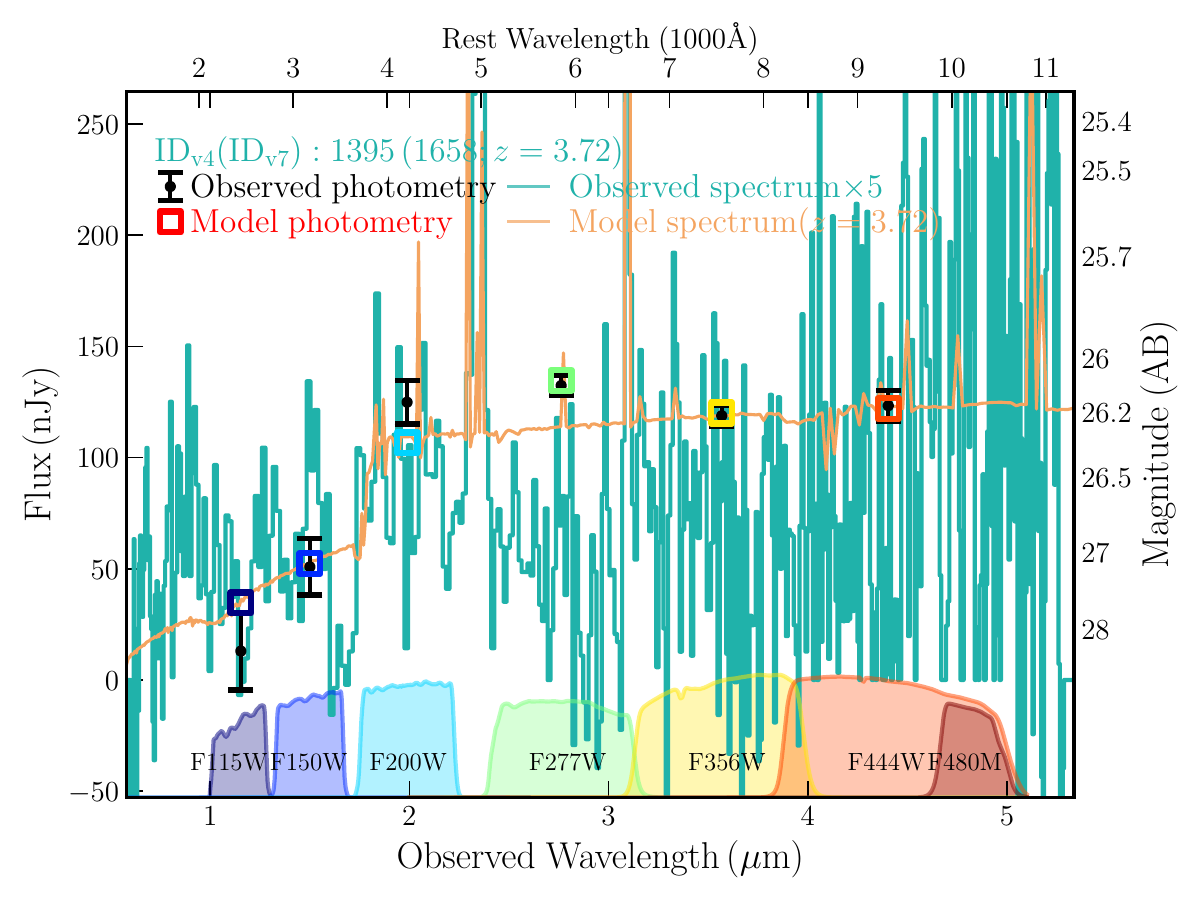}

\begin{minipage}[b]{0.12\hsize}
  \raggedright
  v4 ID 3533\\
  photo-$z$ $\sim$ 11.4\\
  v7 ID 4870\\
  $z = 3.23$
  \includegraphics[width=\hsize]{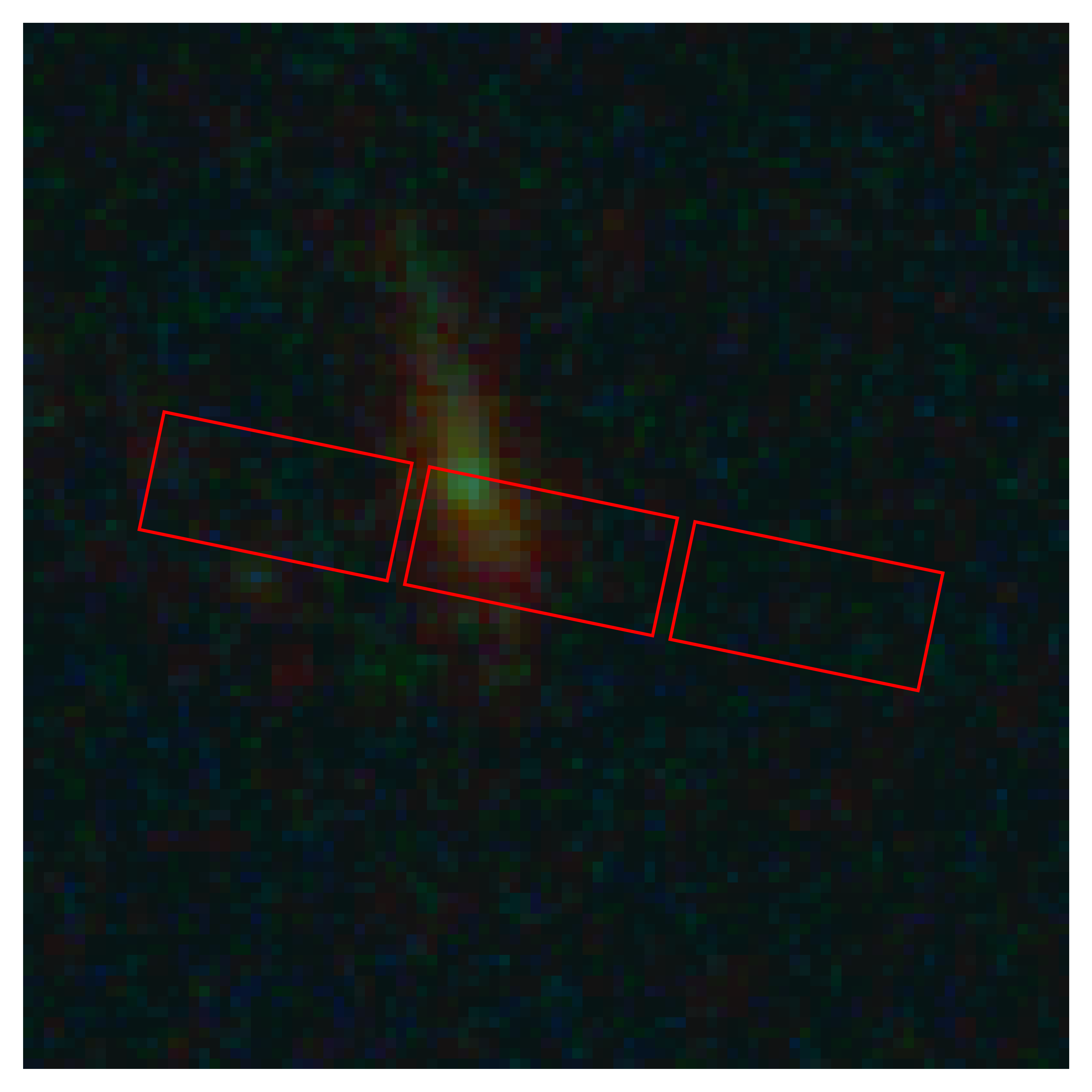}
\end{minipage}
\includegraphics[width=0.50\hsize]{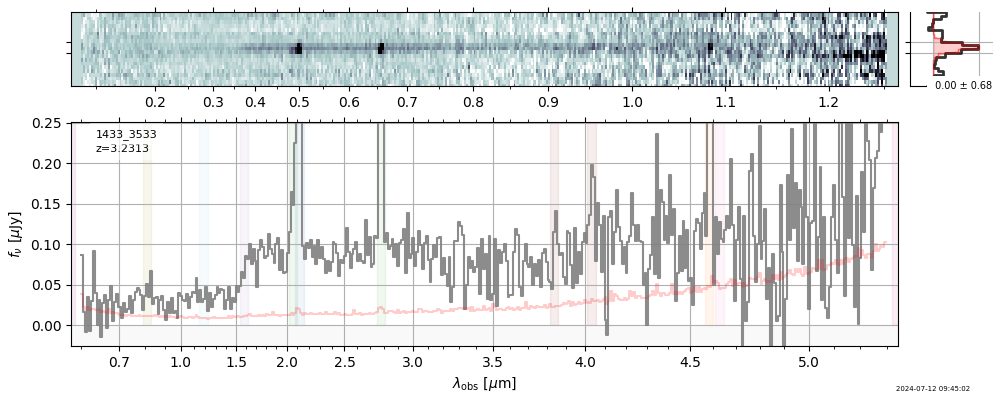}
\includegraphics[width=0.37\hsize, height=1.5in]{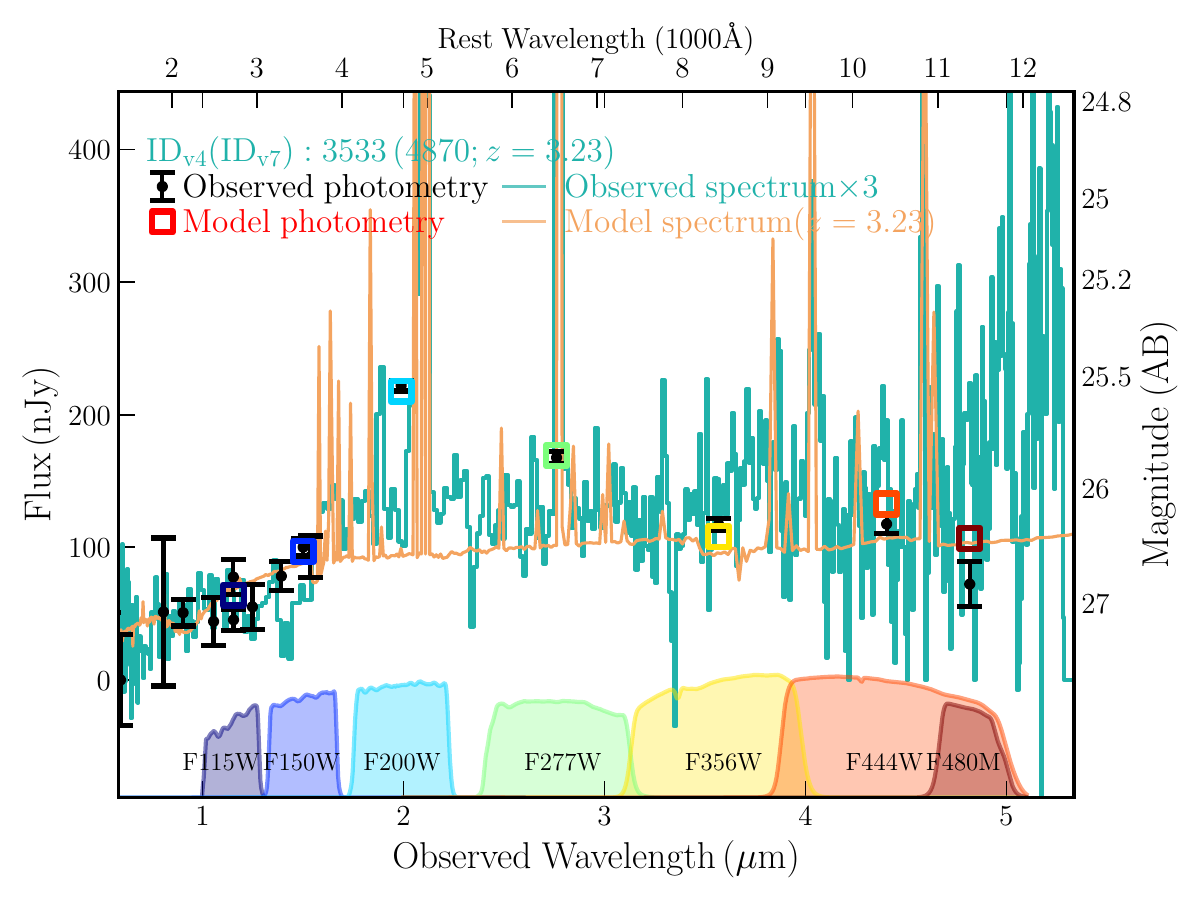}

\caption{
Candidate $z > 11$ galaxies (based on our NIRCam v4 photometry) shown to be interlopers at $z < 4$ with NIRSpec PRISM spectroscopy. In both cases, Balmer breaks are mistaken for Lyman breaks. \textbf{Top:} In the case of v4 3533 (v7 4870), strong emission lines \OIII\ and \Ha\ conspire with the $z = 3.23$ Balmer break to mimic a blue SED ($\beta < -2$) with a Lyman break at $z \sim 11$. \textbf{Bottom:} Emission lines do not significantly affect the photometry of v4 1395 (v7 1658). A faint detection in F150W blueward of the break was missed in the v4 photometric catalog, then detected in the v7 catalog.
}
\label{fig:prism_interlopers}
\end{figure*}

Notably, we also find two high-redshift candidates ($z > 11$) turn out to be low-redshift interlopers ($z < 4$). The high-redshift estimates were based on SED fitting to the NIRCam DJA v4 catalog: $z \sim 12.3$ for v4 1395 (v7 1658) and  $z \sim 11.4$ for v4 3533 (v7 4870). However, NIRSpec spectroscopy revealed significantly lower redshifts ($z_\mathrm{spec}$ = 3.72 and 3.23) as shown in Fig.~\ref{fig:prism_interlopers}. Both galaxies have sharp Balmer breaks that were mistaken for high-redshift Lyman breaks, a classic degeneracy in photometric redshifts.

\JWST\ photometry is often sufficiently precise to resolve such degeneracies. In fact, we found that updating from version 4 to version 7 NIRCam image reductions improved the photometric data, shifting the probability distribution to favor the lower redshift solutions. This illustrates how data reduction pipelines can influence redshift estimates as noted in~\citet{Bagley2023}. The full catalog, including both photometric and spectroscopic redshifts for all candidates, is provided in Appendix~\ref{sec:appendix}.

\section{Conclusion} 
\label{sec:conclusion}
In this paper, we present a new catalog derived from JWST NIRCam and NIRSpec observations of the galaxy cluster MACS0647, revealing a diverse sample of 57 high-redshift galaxy candidates with $z>6$. Spectroscopic observations with NIRSpec confirm redshifts for 14 of these candidates, offering a detailed look into galaxy assembly within the first billion years of cosmic history through their physical properties, from stellar masses and ages to star formation rates and dust content.

The spectroscopic confirmation of these high-redshift galaxies aligns with recent discoveries from other JWST programs, such as CEERS and JADES, which have identified and confirmed galaxies at redshifts $z>10$~\citep{Finkelstein2023, deugenio2024jadesdatarelease3}. These collective efforts underscore the remarkable capability of JWST to probe the distant universe and enhance our understanding of the Epoch of Reionization.

The physical properties derived from our sample indicate a range of stellar masses and star formation rates, suggesting a wide-ranging population of early galaxies. This diversity is consistent with findings from the JADES program, which reported a variety of galaxy sizes and luminosities at similar redshifts~\citep{deugenio2024jadesdatarelease3}. Such variations imply different evolutionary pathways and formation histories, offering valuable insights into the processes governing early galaxy assembly.

Notably, our results, alongside those from CEERS and other surveys, point to the prevalence of relatively low dust content in the majority of early-universe galaxies~\citep{UNCOVER_2024, JADES_Goods_2024, JWST_PEARLS_2023}. This suggests that significant dust production had yet to occur in many of these galaxies, which is consistent with the limited time available for dust formation and enrichment processes in the early universe.

In summary, our work expands the growing sample of spectroscopically confirmed high-redshift galaxies and reinforces the emerging picture of a diverse early galaxy population. The range of physical properties observed in MACS0647 highlights the complexity of galaxy assembly within the first billion years. Continued spectroscopic follow-up with JWST will be essential for refining these results and for connecting early galaxy populations to the progress of reionization.


\section{Acknowledgments}
K.W. thanks the William H. Miller Graduate Fellowship and the MDSGC Observatory Fellowship. K.W. also thanks the LSST-DA Data Science Fellowship Program, which is funded by LSST-DA, the Brinson Foundation, and the Moore Foundation; his participation in the program has benefited this work.

This work is based on observations made with the NASA/ESA/CSA \JWST.
The \JWST\ data presented in this article were obtained from the Mikulski Archive for Space Telescopes (MAST) at the Space Telescope Science Institute. 
The Association of Universities for Research in Astronomy (AURA), Inc.~operates MAST under NASA contract NAS 5-03127 for \JWST.

This work was supported by JSPS KAKENHI Grant Numbers JP25H00662, JP22K21349.

%

\vspace{5mm}


\software{
\grizli\ \citep{Brammer2022},
astropy \citep{astropysoftware1,astropysoftware2,astropysoftware3},
matplotlib \citep{Hunter:2007},
NumPy \citep{harris2020array},
SciPy \citep{2020SciPy-NMeth},
STScI \textit{JWST} Calibration Pipeline (\url{jwst-pipeline.readthedocs.io}; \citealt{Rigby2022}, \citealt{Bushouse_2023}),
\msaexp\ \citep{Brammer_msaexp},
PyNeb \citep{Luridiana_2014}.
}



%


\clearpage
\bibliography{papers}{}
\bibliographystyle{aasjournal}



\section{Appendix} 
\label{sec:appendix}
\subsection{Full Candidate List}
\begin{table*}[htbp]
\centering
\setlength{\tabcolsep}{8pt}
\caption{\label{tab:catalog2}
High-redshift candidates identified in this work, including cross-matches with \citet{Bradley2014} and \citet{Meena2023}. 
Magnification factors are derived from the MACS0647 mass model of \citet{Nishida2025}, constructed with \textsc{glafic} \citep{Oguri2010, Oguri2021}, with 68\% confidence intervals quoted. 
\textbf{Note:} Little Red Dots (LRDs) are marked with an asterisk ($^{*}$); their SED fits may be unreliable due to extreme colors or AGN contamination.
}
\begin{tabular}{lccccccccc}
\hline
ID(v7) & ID(v4) & Bradley 14 ID & Meena 24 ID & RA & DEC & F200W mag & $z_{\rm phot}$ & $z_{\rm spec}$ & $\mu$ \\
\hline
173 & 116 & \nodata & \nodata & 101.964896 & 70.184156 & 27.422 & 6.71 & \nodata & $1.10^{+0.01}_{-0.01}$ \\
274 & 198 & \nodata & \nodata & 101.935274 & 70.185656 & 27.744 & 9.52 & \nodata & $1.06^{+0.01}_{-0.01}$ \\
587 & 486 & \nodata & \nodata & 101.868833 & 70.189611 & 27.996 & 9.64 & \nodata & $1.02^{+0.01}_{-0.01}$ \\
792 & 661 & \nodata & \nodata & 101.978383 & 70.19233 & 28.094 & 6.69 & \nodata & $1.14^{+0.01}_{-0.01}$ \\
904 & 753 & \nodata & \nodata & 101.87733 & 70.193708 & 29.455 & 7.15 & \nodata & $1.03^{+0.01}_{-0.01}$ \\
1021 & 852 & \nodata & \nodata & 101.96016 & 70.195471 & 29.895 & 6.70 & \nodata & $1.12^{+0.01}_{-0.01}$ \\
1107 & 926 & \nodata & \nodata & 101.957695 & 70.196484 & 26.004 & 6.72 & \nodata & $1.12^{+0.01}_{-0.01}$ \\
1133 & 946 & \nodata & \nodata & 101.950988 & 70.196883 & 29.287 & 6.83 & \nodata & $1.11^{+0.01}_{-0.01}$ \\
1175 & 986 & \nodata & \nodata & 101.975962 & 70.197477 & 28.262 & 9.61 & \nodata & $1.17^{+0.01}_{-0.02}$ \\
1352 & 1139 & \nodata & \nodata & 101.906024 & 70.19979 & 28.262 & 6.86 & \nodata & $1.06^{+0.01}_{-0.01}$ \\
1471 & 1243 & \nodata & \nodata & 101.924964 & 70.201544 & 27.110 & 9.16 & \nodata & $1.08^{+0.02}_{-0.01}$ \\
1626 & 1378 & \nodata & \nodata & 101.961021 & 70.203511 & 28.324 & 6.20 & \nodata & $1.16^{+0.01}_{-0.02}$ \\
1658 & 1395 & \nodata & \nodata & 101.903143 & 70.203745 & 26.019 & 12.49 & 3.72 & $1.06^{+0.01}_{-0.01}$ \\
1857 & 1559 & \nodata & \nodata & 101.932349 & 70.206767 & 27.863 & 10.96 & \nodata & $1.12^{+0.02}_{-0.02}$ \\
2120 & 1714 & \nodata & \nodata & 101.983563 & 70.209769 & 26.919 & 6.71 & \nodata & $1.26^{+0.02}_{-0.02}$ \\
2121 & 1715 & \nodata & \nodata & 101.983171 & 70.209787 & 27.316 & 6.69 & 6.14 & $1.26^{+0.02}_{-0.02}$ \\
2193$^{*}$ & 1742 & \nodata & \nodata & 101.892498 & 70.210177 & 26.400 & 7.54 & 5.81 & $1.10^{+0.02}_{-0.01}$ \\
2172 & 1751 & \nodata & \nodata & 101.971957 & 70.21037 & 27.197 & 9.00 & \nodata & $1.25^{+0.02}_{-0.02}$ \\
2426 & 1882 & \nodata & \nodata & 101.944994 & 70.212959 & 29.287 & 6.82 & \nodata & $1.19^{+0.02}_{-0.02}$ \\
2510 & 1912 & \nodata & \nodata & 101.977783 & 70.213676 & 27.291 & 6.69 & \nodata & $1.30^{+0.02}_{-0.02}$ \\
2551 & 1924 & \nodata & \nodata & 101.902034 & 70.213857 & 25.843 & 9.58 & \nodata & $1.14^{+0.02}_{-0.02}$ \\
2619 & 1944 & \nodata & \nodata & 101.986107 & 70.214346 & 25.715 & 6.12 & \nodata & $1.32^{+0.02}_{-0.02}$ \\
2846 & 2019 & \nodata & \nodata & 101.973309 & 70.215423 & 27.744 & 6.82 & \nodata & $1.31^{+0.02}_{-0.02}$ \\
4010 & 2876 & 2175 & \nodata & 101.984019 & 70.229488 & 27.395 & 6.67 & 6.10 & $1.80^{+0.04}_{-0.04}$ \\
4213 & 3035 & \nodata & \nodata & 101.922863 & 70.234139 & 28.094 & 8.93 & \nodata & $1.78^{+0.06}_{-0.05}$ \\
4268 & 3090 & 1976 & \nodata & 101.954327 & 70.235365 & 26.444 & 6.31 & 6.13 & $1.98^{+0.07}_{-0.06}$ \\
4336 & 3138 & \nodata & \nodata & 101.906778 & 70.236517 & 27.175 & 7.03 & \nodata & $1.81^{+0.06}_{-0.05}$ \\
4393 & 3186 & \nodata & \nodata & 101.971685 & 70.237378 & 28.797 & 8.90 & \nodata & $3.37^{+0.13}_{-0.14}$ \\
4411 & 3208 & \nodata & \nodata & 101.918791 & 70.237614 & 26.422 & 6.58 & 6.12 & $2.03^{+0.07}_{-0.07}$ \\
4533 & 3308 & \nodata & 16.4 & 101.952449 & 70.239294 & 27.395 & 6.70 & 6.13 &  $2.77^{+0.13}_{-0.11}$ \\
4797 & 3493 & \nodata & \nodata & 101.920053 & 70.241989 & 27.707 & 10.02 & \nodata 
& $3.51^{+0.18}_{-0.17}$ \\
4824 & 3504 & \nodata & \nodata & 101.903249 & 70.242103 & 29.895 & 8.36 & \nodata & $2.42^{+0.08}_{-0.08}$ \\
4870 & 3533 & \nodata & 15.1 & 101.999421 & 70.242449 & 25.515 & 11.41 & 3.23 & $6.58^{+0.45}_{-0.37}$ \\
4919 & 3567 & \nodata & 5.2 & 101.9216 & 70.242932 & 27.175 & 6.91 & \nodata & $4.01^{+0.20}_{-0.20}$ \\
4922 & 3568 & \nodata & \nodata & 101.90398 & 70.242978 & 27.782 & 9.57 & 9.25 & $2.72^{+0.09}_{-0.09}$ \\
5091 & 3643 & \nodata & \nodata & 101.97757	& 70.244199 & 27.707 & 6.10 & \nodata & $5.01^{+0.27}_{-0.27}$ \\
5191 & 3754 & 1411 & 10.2 & 101.920529 & 70.244899 & 25.605 & 7.33 & 7.47 & $7.51^{+0.44}_{-0.43}$ \\
5595 & 3983 & \nodata & 16.2 & 101.944871 & 70.248877 & 22.249 & 6.69 & \nodata & $10.97^{+3.67}_{-2.10}$ \\
5591 & 3989 & \nodata & \nodata & 101.940754 & 70.249101 & 28.262 & 6.30 & 6.14 & $1.99^{+0.10}_{-0.10}$ \\
5594 & 3993 & \nodata & 10.1 & 101.919592 & 70.24904 & 26.193 & 7.34 & \nodata & $6.78^{+0.40}_{-0.40}$ \\
\end{tabular}
\end{table*}
\FloatBarrier
\begin{table*}[htbp]  
\centering
\setlength{\tabcolsep}{8pt}
\caption{\label{tab:catalog2b}
High-redshift candidates identified in this work (Part~2), including cross-matches with \citet{Bradley2014} and \citet{Meena2023}. 
Magnification values follow the same lens model described in Table~\ref{tab:catalog2}.
}
\begin{tabular}{lccccccccc}
\hline
ID(v7) & ID(v4) & Bradley 14 ID & Meena 24 ID & RA & DEC & F200W mag & $z_{\rm phot}$ & $z_{\rm spec}$ & $\mu$ \\
\hline
5881 & 4186 & \nodata & 5.1 & 101.921109 & 70.251551 & 26.772 & 6.33 & \nodata & $4.60^{+0.24}_{-0.22}$ \\
5931 & 4219 & \nodata & \nodata & 101.916697 & 70.252253 & 29.455 & 6.66 & 6.12 & $7.41^{+0.63}_{-0.57}$ \\
6339 & 4514 & \nodata & \nodata & 101.99504 & 70.255751 & 29.014 & 6.81 & \nodata & $2.92^{+0.10}_{-0.09}$ \\
6736 & 4775 & \nodata & \nodata & 101.934471 & 70.259417 & 27.131 & 6.52 & \nodata & $6.78^{+0.19}_{-0.16}$ \\
6839 & 4838 & \nodata & \nodata & 101.934269 & 70.260138 & 28.147 & 6.30 & \nodata & $3.74^{+0.19}_{-0.16}$ \\
6871 & 4865 & \nodata & \nodata & 102.010301 & 70.260557 & 29.142 & 8.93 & \nodata & $1.87^{+0.05}_{-0.04}$ \\
6901 & 4893 & \nodata & \nodata & 101.965589 & 70.26072 & 28.147 & 6.56 & \nodata & $2.11^{+0.10}_{-0.09}$ \\
6941 & 4914 & \nodata & \nodata & 101.964944 & 70.260979 & 28.900 & 6.63 & \nodata & $2.09^{+0.09}_{-0.09}$ \\
6943 & 4915 & \nodata & \nodata & 101.945294 & 70.261003 & 28.203 & 9.55 & \nodata & $2.93^{+0.15}_{-0.14}$ \\
6952 & 4920 & \nodata & \nodata & 101.964462 & 70.261078 & 28.324 & 7.03 & \nodata & $2.10^{+0.09}_{-0.09}$ \\
7120 & 5010 & 339 & \nodata & 101.924688 & 70.261861 & 25.534 & 6.16 & 6.11 & $2.63^{+0.09}_{-0.07}$ \\
7193 & 5057 & \nodata & 16.3 & 101.966915 & 70.262567 & 29.653 & 6.69 & \nodata & $1.89^{+0.08}_{-0.07}$ \\
7282 & 5094 & \nodata & \nodata & 101.925405 & 70.263143 & 26.467 & 7.37 & 7.00 & $2.32^{+0.07}_{-0.06}$ \\
7397 & 5156 & \nodata & \nodata & 101.972484 & 70.264076 & 29.142 & 9.38 & \nodata & $1.68^{+0.06}_{-0.06}$ \\
7576 & 5332 & \nodata & \nodata & 101.963323 & 70.265634 & 28.324 & 7.28 & \nodata & $1.60^{+0.05}_{-0.05}$ \\
7604 & 5352 & \nodata & \nodata & 101.98205 & 70.26602 & 27.509 & 6.70 & \nodata & $1.51^{+0.05}_{-0.04}$ \\
7697 & 5420 & \nodata & \nodata & 101.922465 & 70.266946 & 27.672 & 6.56 & \nodata & $1.84^{+0.05}_{-0.04}$ \\
\hline
\end{tabular}
\end{table*}


\clearpage
\subsection{NIRSpec Fluxes and Spectra}
\begin{table*}[htbp]
\centering
\caption{Emission line fluxes (1e-20 cgs) measured for galaxies observed with NIRSpec PRISM.} \label{tab:reversed_emission_lines}
\begin{tabular}{lccccccccc}
\hline
ID(v7) & 1658 & 2121 & 2193 & 4411 & 4533 & 4870 & 4922 & 5191 & 5591 \\
\hline
OII & 89$\pm$21 & 26$\pm$8 & \nodata & 91$\pm$14 & 110$\pm$8 & \nodata & \nodata & 21$\pm$5 & \nodata \\
HeI-3889 & 87$\pm$19 & \nodata & \nodata & \nodata & 36$\pm$9 & \nodata & \nodata & 23$\pm$6 & \nodata \\
Ha+NII & 150$\pm$7 & 106$\pm$7 & 994$\pm$16 & 123$\pm$9 & 454$\pm$9 & 245$\pm$10 & \nodata & \nodata & 43$\pm$5 \\
SII & 18$\pm$5 & \nodata & \nodata & \nodata & 21$\pm$4 & \nodata & \nodata & \nodata & \nodata \\
HeI-7065 & 16$\pm$5 & \nodata & \nodata & \nodata & \nodata & \nodata & \nodata & \nodata & 13$\pm$4 \\
SIII-9531 & 17$\pm$5 & \nodata & \nodata & \nodata & \nodata & 25$\pm$6 & \nodata & \nodata & \nodata \\
OIII-1663 & \nodata & 155$\pm$40 & \nodata & \nodata & \nodata & \nodata & \nodata & \nodata & \nodata \\
OIII-4363 & \nodata & 26$\pm$6 & \nodata & \nodata & 27$\pm$6 & \nodata & \nodata & 14$\pm$4 & \nodata \\
Hb & \nodata & 34$\pm$5 & 214$\pm$10 & 47$\pm$8 & 206$\pm$7 & 87$\pm$15 & 18$\pm$5 & 70$\pm$5 & \nodata \\
OIII-4959 & \nodata & 54$\pm$5 & 336$\pm$11 & 82$\pm$8 & 436$\pm$10 & \nodata & 34$\pm$6 & 103$\pm$5 & 23$\pm$4 \\
OIII-5007 & 414$\pm$15 & 184$\pm$7 & 908$\pm$18 & 241$\pm$10 & 1250$\pm$18 & 671$\pm$22  & 65$\pm$6 & 334$\pm$8 & 72$\pm$5 \\
HeI-5877 & \nodata & \nodata & 37$\pm$6 & \nodata & 20$\pm$4 & \nodata & \nodata & \nodata & \nodata \\
OII-7325 & \nodata & \nodata & \nodata & 31$\pm$10 & \nodata & \nodata & \nodata & \nodata & \nodata \\
NeIII-3867 & \nodata & \nodata & \nodata & 50$\pm$16 & 56$\pm$9 & \nodata & \nodata & 22$\pm$6 & \nodata \\
NeIII-3968 & \nodata & \nodata & \nodata & \nodata & 52$\pm$7 & \nodata & \nodata & 24$\pm$5 & \nodata \\
Hd & \nodata & \nodata & \nodata & \nodata & 64$\pm$6 & \nodata & \nodata & 28$\pm$6 & \nodata \\
Hg & \nodata & \nodata & \nodata & \nodata & 100$\pm$7 & \nodata & \nodata & 24$\pm$4 & \nodata \\
HeI-1083 & \nodata & \nodata & \nodata & \nodata & \nodata &  44$\pm$4  & \nodata & \nodata & \nodata \\
NIII-1750 & \nodata & \nodata & \nodata & \nodata & \nodata & \nodata & 88$\pm$27 & \nodata & \nodata \\
\hline
12+log(O/H) & & $7.58\pm0.27$ & & $8.14\pm0.08$ & $7.82\pm0.10$ & & & $7.50\pm0.18$ & \\
12+log(O/H)$^{a}$ & $\sim$8.00$^{b}$ & $7.61\pm0.16$ & $7.43\pm0.03$ & $7.57\pm0.17$ & $7.74\pm0.05$ & $\sim$8.00$^{b}$ & $7.33\pm0.21$ & $7.51\pm0.06$ & $>7.67^{c}$ \\
log(U) & & $-2.02\pm0.06$ & & $-2.41\pm0.07$ & $-1.89\pm0.13$ & & & $-1.79\pm0.15$ & \\
\hline
\end{tabular}
\tablecomments{
$^{a}$ Derived from strong lines (i.e., R3 = OIII / H$\beta$) adopting the relation from \citet{Nakajima2023}. 
$^{b}$ R3 is higher than the peak of the relation.
$^{c}$ Lower limit from 3$\sigma$ of H$\beta$.
}
\end{table*}  
\label{tab:emission_lines}

\begin{deluxetable*}{ccccccccccc}
\setlength{\tabcolsep}{4pt}
\tablecaption{Emission line fluxes (1e-20 cgs) and FWHM (km/s) measured for galaxies observed with NIRSpec high-res G395H.} \label{tab:spec_g395h}
\tablewidth{\columnwidth}
\tablehead{
\colhead{ID} & 
\colhead{ID} & 
\colhead{$z$} & 
\colhead{\Hb} & 
\colhead{\Hb} & 
\colhead{[OIII]} & 
\colhead{[OIII]} & 
\colhead{[OIII]} & 
\colhead{[OIII]} & 
\colhead{\Ha} & 
\colhead{\Ha} 
\\
\colhead{v7} & 
\colhead{v4} & 
\colhead{} &
\colhead{4863} & 
\colhead{4863} & 
\colhead{4960} & 
\colhead{4960} & 
\colhead{5008} & 
\colhead{5008} & 
\colhead{6565} & 
\colhead{6565} 
\\
\colhead{} & 
\colhead{} & 
\colhead{} &
\colhead{flux} & 
\colhead{FWHM} & 
\colhead{flux} & 
\colhead{FWHM} & 
\colhead{flux} & 
\colhead{FWHM} & 
\colhead{flux} & 
\colhead{FWHM}
}
\startdata
5191 & 3754 & 7.466 &
$231 \pm 60$ & $152^{+63}_{-42}$ & 
$335 \pm 59$ & $143^{+35}_{-27}$ & 
$1100 \pm 79$ & $128^{+12}_{-11}$ & 
\nodata & \nodata 
\\
7282 & 5094 & 6.999 &
\nodata & \nodata & 
$388 \pm 23$ & $202^{+14}_{-13}$ & 
$1067 \pm 38$ & $182^{+50}_{-48}$ & 
\nodata & \nodata 
\\
4268 & 3090 & 6.127 &
$97\pm 13$  & $157^{+28}_{-24}$ & 
$197\pm 9$ & $158 \pm 8$ & 
$638\pm 19$ & $157 \pm 5$ & 
$366\pm 38$ & $155^{+20}_{-17}$ 
\\
5931 & 4219 & 6.117 &
$29 \pm 7$ & $104^{+35}_{-34}$ & 
$43 \pm 2$ & $102 \pm 5$ & 
$180 \pm 10$ & $126^{+8}_{-7}$ & 
$90 \pm 24$ & $105^{+43}_{-29}$ 
\\
7120 & 5010 & 6.115 &
$1670 \pm 168$ & $213^{+23}_{-22}$ & 
$3842 \pm 231$ & $193 \pm 8$ & 
$12460 \pm 758$ & $197 \pm 9$ & 
$4881 \pm 301$ & $172^{+11}_{-10}$ 
\\
4010 & 2876 & 6.101 &
$43 \pm 5$ & $153^{+22}_{-19}$ & 
$76 \pm 4$ & $166 \pm 8$ & 
$240 \pm 11$ & $172 \pm 8$ & 
$160 \pm 33$ & $161^{+48}_{-33}$ 
\\
2193 & 1742 & 5.797 &
$133 \pm 15$ & $152 \pm 16$ & 
$373 \pm 30$ & $148^{+11}_{-10}$ & 
$1059 \pm 56$ & $160^{+7}_{-6}$ & 
\nodata & \nodata 
\enddata
\end{deluxetable*}  

\vspace{-5mm}

\begin{figure*} 
\includegraphics[height=3.5cm,keepaspectratio]{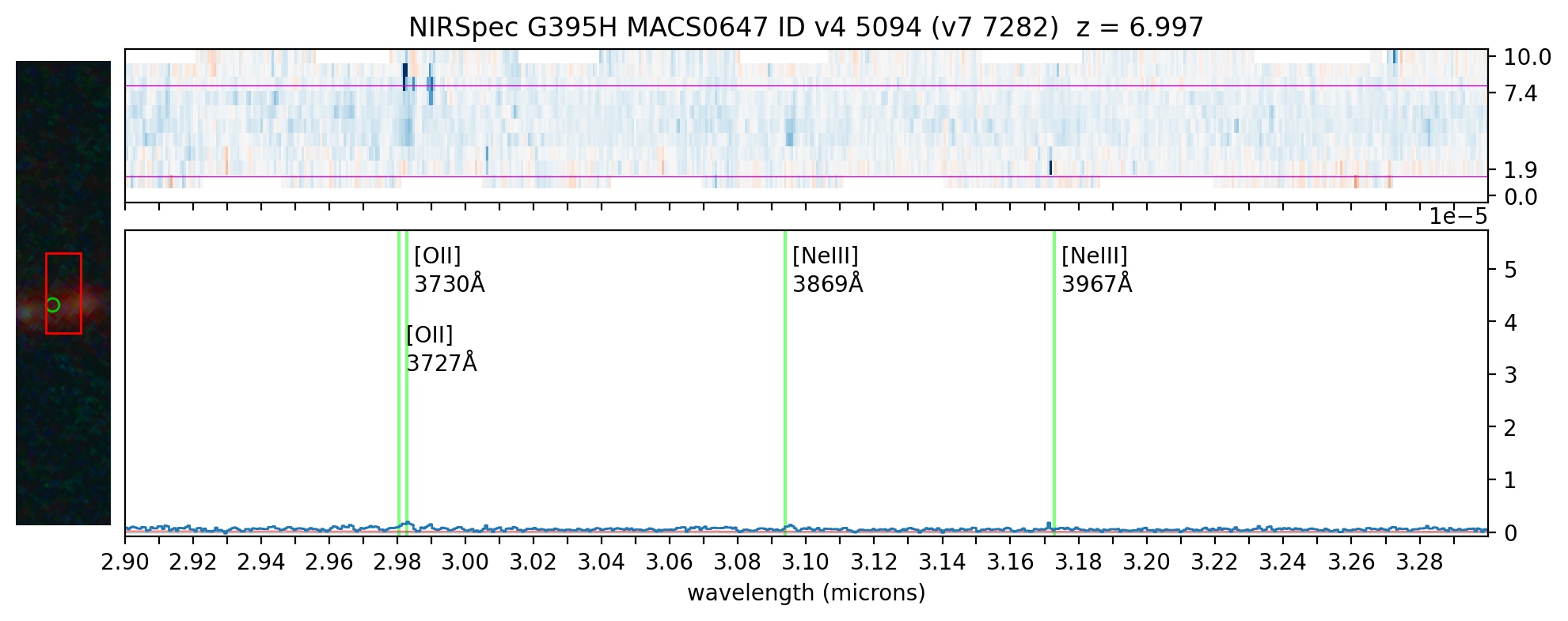}
\includegraphics[height=3.5cm,keepaspectratio]{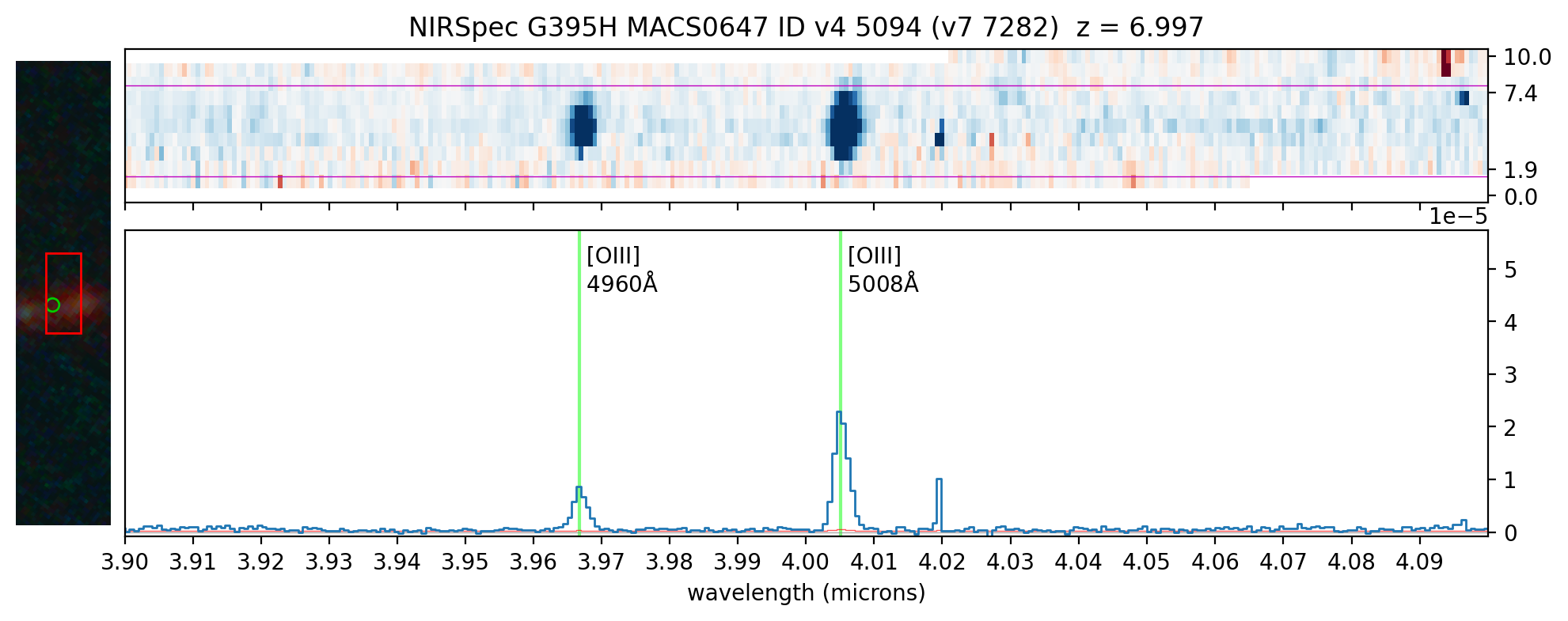}
\caption{
NIRSpec G395H spectrum of galaxy v4 5494 (v7 7282) at $z = 6.997$, with clear detections of \Hb, \OIII, and \Ha. The strong \OIII\ emission indicates a highly ionized, low-metallicity ISM typical of $z\sim7$ galaxies.
}
\label{fig:5494_g395h}
\end{figure*}

\end{document}